\documentclass[twocolumn,aps,floats,floatfix,nofootinbib,prd,superscriptaddress,tightenlines]{revtex4}
\usepackage{axodraw}
\usepackage{graphicx}
\usepackage{bm}
\usepackage{pstricks}

\begin{document}

\title{An Effective Field Theory of Gravity for Extended Objects}

\author{Walter D. Goldberger}
\affiliation{Department of Physics, Yale University, New Haven,  CT 06511}

\author{Ira Z. Rothstein}
\affiliation{Department of Physics, Carnegie Mellon University, Pittsburgh, PA 15213}

\date{March 2005}

\begin{abstract}

Using Effective Field Theory (EFT) methods we present a Lagrangian formalism which describes the dynamics of non-relativistic extended objects coupled to gravity.  The formalism is  relevant to understanding the gravitational radiation power spectra emitted by binary star systems, an important class of candidate signals for gravitational wave observatories such as LIGO or VIRGO.  The EFT allows for a clean separation  of the three relevant scales: $r_s$, the size of the compact objects, $r$ the orbital radius and $r/v$, the wavelength of the physical radiation (where the velocity $v$ is the expansion parameter).  In the EFT radiation is systematically included in the $v$ expansion without need to separate integrals into near zones and radiation zones.  Using the EFT, we show that the renormalization of ultraviolet divergences which arise at $v^6$ in post-Newtonian (PN) calculations requires the presence of two non-minimal worldline gravitational couplings linear in the Ricci curvature.  However, these operators can be removed by a redefinition of the metric tensor, so that the divergences at arising at $v^6$ have no physically observable effect.  Because in the EFT finite size features are encoded in the coefficients of non-minimal couplings, this implies a simple proof of the decoupling of internal structure for spinless objects to at least order $v^6$.  Neglecting absorptive effects, we find that the power counting rules of the EFT indicate that the next set of short distance operators, which are quadratic in the curvature and are associated with tidal deformations, do not play a role until order $v^{10}$.  These operators, which encapsulate finite size properties of the sources, have coefficients that can be fixed by a matching calculation.  By including the most general set of such operators, the EFT allows one to work within a point particle theory to arbitrary orders in $v$.  \end{abstract}

\maketitle

\section{Introduction}

The experimental progam in gravitational wave detection that is currently under way~\cite{LIGO,VIRGO} has brought renewed attention to the problem of obtaining high accuracy predictions for the evolution and gravitational radiation power spectra of binary star systems with neutron star or black hole  constituents.  The conventional approach to calculating the initial inspiral of the system as it slowly loses energy to gravitational radiation employs the post-Newtonian (PN) approximation to general relativity.  Essentially, this formalism consists of systematically solving the Einstein equations with non-relativistic (NR) sources as a power series expansion in  $v<1$, where $v$ is a typical three-velocity of the system under consideration~\cite{EIH,PNformalism}.  For binary systems that can be detected by LIGO, which have binary constituents with masses in the range $\sim (1-10)\, m_\odot$, the inspiral phase spans roughly $0.1<v<0.4$, corresponding to orbital separations from about $100$ to a few times $(m/m_\odot)\, \mbox{km}$.

A detector such as LIGO is particularly sensitive to the time varying phase $\Delta\phi(t)$ of the gravitational waves emitted during the evolution of the binary system~\cite{dataanalysis}.  In the PN formalism, this observable can be obtained by calculating the mechanical energy $E$ as well as the total power $P$ emitted in gravitational radiation as functions of the orbital parameters of the binary system.  Taking for illustration the extreme mass ratio limit and using conservation of energy, $dE/dt = - P$, for circular orbits the gravitational wave phase seen by the detector is then given by 
\begin{equation}
\label{eq:phase}
\Delta \phi(t) = \int_{t_0}^t d\tau \omega(\tau) = {2\over G_N M}\int^{v_0} _{v(t)}d v' v'^3 {dE/dv'\over P(v')},
\end{equation}  
where $\omega(\tau)$ is the wave's frequency and $M$ the mass of the heavy component star.  Because a typical inspiral event will sweep a large number of radians of phase as it scans the detector frequency band (e.g, for LIGO, with a frequency range $1-10^4$~Hz,  the number of cycles for a neutron star/neutron star binary is of order $10^4$), to compare the predictions of GR with the experimental data requires computing the observable $\Delta\phi(t)$ to extremely high order in the velocity expansion.  For instance, for binary neutron star inspirals in the LIGO band, one must know $\Delta\phi(t)$ to ${\cal O}(v^6)$ (or ``3PN") beyond the leading order quadrupole radiation results, and higher order terms in $v$ may be necessary for tracking the phase variation of the gravitational waves emitted by more massive objects~\cite{dataanalysis}.   

Several independent research groups have applied variants of the PN formalism to the computation of  high order corrections to the gravitational wave observables during the binary inspiral phase~\cite{PNv3rad,PNv4pot,PNv4rad,PNv5pot,PNv56rad,PNv6pot}.  Heroic calculational efforts have yielded in the last few years the $v^6$ analytic results necessary to fully interpret the LIGO data~\cite{PNv56rad,PNv6pot}.  While the independent calculations are in agreement including terms up to $v^5$ (or ``$2.5$PN") beyond the predictions of Newtonian gravity, the $v^6$ corrections do not give well defined results.   At this order, there seems to be an ambiguity in the PN expansions for the binary system, stemming from scheme dependence in the regularization of singular integrals that arise in the perturbative series.  Recently, refs.~\cite{PNambiguity,PNdimreg} have proposed a resolution of this ambiguity purely within the framework of the PN expansion. 

The singular integrals that cause trouble at high orders in $v$ can be seen to arise from treating the binary star constituents as point particle sources coupled to general relativity.  It is then clear that these singularities are simply the usual ultraviolet (UV) divergences that appear in any field theory coupled to point sources.  Their presence is merely a sign that the formalism has limited predictive power and needs to be supplemented by a more complete model that accounts for the fact that the sources are not really pointlike.  While in principle one could obtain finite, unambiguous results for experimental observables in this more complete theory, the price would be much greater computational complexity.  Even then, it would be difficult to disentangle the effects that depend upon short distance physics from those that are solely a consequence of the gravitational dynamics.  Since the goal of the gravitational wave experiments is both to probe the internal structure of strongly gravitating astrophysical systems as well as to test the dynamics of general relativity (or possible deviations from it), it is critical to have a computational approach that can cleanly separate these two aspects of the binary problem.

In this paper, we propose a Lagrangian framework for systematically calculating to any order in the PN expansion.  Within our theory the ``ambiguities"  that plague the conventional PN techniques at order $v^6$ and beyond have a natural place in the renormalization program.  Our formalism is based on effective field theory (EFT) reasoning, and it is motivated by the EFT approach to NR bound state problems in quantum field theories such as QED and QCD~\cite{NREFT,LMR}.  Rather than trying to resolve the point particle singularities by resorting to a specific model of the short distance physics, in an EFT framework~\cite{EFTrev} we systematically parametrize our ignorance of this structure by including in the effective point particle Lagrangian the most general set of operators consistent with the symmetries of the well understood, long wavelength physics.  In our case, this long wavelength physics is general relativity, and the symmetry that constrains the dynamics is just general coordinate invariance.  

The operators in the point particle action have coefficients which encapsulate the properties of the structure of the extended sources.   UV divergences encountered in calculations in the EFT can then be dealt with using any convenient regulator, and renormalized by adjusting the operator coefficients\footnote{In this paper we employ dimensional regularization, and modified minimal subtraction (the $\overline{MS}$ scheme).  This has the advantage of not introducing spurious length scales that could spoil dimensional analysis arguments within the EFT. See~\cite{PNdimreg} for a discussion of dimensional regularization in the context of conventional PN formalisms}.  Given a model of stellar structure, the precise value of these coefficients can be obtained by a UV matching calculation that we describe below in Sec.~\ref{sec:matching}.  Note that because these operators are intrinsic to the structure of the sources, the matching procedure can be carried out for a single source in isolation, independently of the complicated binary star dynamics.  In this way the EFT manages to disentangle the model dependent aspects of the gravitational wave signals from the features that arise as unambiguous predictions of general relativity.  

An EFT also has the advantage of having manifest power counting in the expansion parameter of the theory, in our case the velocity $v$.  This means that in the EFT it is possible to calculate, using simple scaling arguments, the order in $v$ which a given term in the perturbative series (i.e, a typical Feynman diagram obtained from the vertices of the effective Lagrangian or an operator in the effective theory) first contributes to a given physical observable.  In particular, one can determine ahead of any calculation the number of terms in the point particle Lagrangian that will appear at any order in $v$.  

As an example, using our formalism we will see that for binaries with non-spinning black hole or neutron star constituents, the proper time term is sufficient to describe the physics up to $v^5$.   The apearance of logarithmic divergences in the PN expansion that first appear at order $v^6$ indicates that two new non-minimal terms whose form is dictated by general covariance must be included at this order to consistently renormalize the point particle theory.  However, the coefficients of these operators can be shifted arbitrarily by making field redefinitions of the metric, which indicates that the $v^6$ operators cannot contribute to observables.  This gives a clean argument for the absence of $v^6$ finite size effects in the predictions for gravitational wave templates.  Ignoring the effects of absorption\footnote{In the case of extreme mass ratio binaries where the large mass component is a black hole, it is known that absorption into the black hole first appears at order $v^8$~\cite{poisson}.   We will show how absorptive effects are incorporated into the EFT in a future paper.}, we also find that the next set of point particle operators, including for instance operators quadratic in the Riemann curvature, do not contribute until order $v^{10}$.  These operators, which cannot be removed by field redefinitions, are associated with the leading effects of tidal forces exerted on an isolated neutron star or black hole by an external gravitational field.  The fact that such tidal operators do not arise until $v^{10}$ is consistent with purely Newtonian estimates which also conclude that tidal forces are $v^{10}$ effects\footnote{We thank Kip Thorne for pointing this out to us.  See also~\cite{will} for a recent review.}.  

Finally, an EFT has the advantage that it is possible to use renormalization group (RG) methods to understand in a simple way logarithms of scale ratios that may arise in the perturbative series for a given observable.  For example, in the NR expansion of the binary star gravitational wave observables, one encounters logarithmic terms of the form $v^n(\ln\alpha)^m,$ where $\alpha=r_s/r$ is the ratio of a distance scale $r_s$ that characterizes the size of the binary constituents and $r$ is a distance of order the orbital radius (note that for black hole binaries, the virial theorem relates $\alpha\sim v^2$).  We will show that in our EFT, the coefficients of operators in the effective point particle action undergo non-trivial RG flows, so that the logarithms of $\alpha$ (tidal logarithms) can be obtained by RG running from a renormalization scale $\mu\sim 1/r_s$ where the point particle theory is initially defined down to a scale $\mu\sim 1/r$ where the binary dynamics takes place.   Moreover, the complete structure of the RG flows is encoded in the black hole metric.

 As an explicit example of the RG method, we will give a simplified derivation of the terms at ${\cal O}(v^6)$ in the two-body interaction Lagrangian (or equivalently the binary gravitational binding energy) that are enhanced by $\ln\alpha$.  We then show explicitly that the effect of these logarithms do not contribute to physical observables. This result must follow since, as we already mentioned above, the $v^6$ worldline operators can be removed by redefining the metric.   However, this calculation is still useful as a template for determining the logarithmically enhanced terms at higher orders in $v$ that stem from true finite size effects.

 The outline of the paper is as follows.  In the next section we formulate our EFT.   The starting point is a theory of relativistic point particles coupled to gravity.  This theory is appropriate for discussing the dynamics of extended sources at inverse momentum transfers larger than their size $r_s$.  We then describe the general procedure for matching this theory at the mass scale\footnote{We use units $\hbar=c=1$.} $\mu\sim 1/r$, where the binary  orbital dynamics takes place, onto an EFT that is optimized for the NR limit (NRGR).  The resulting EFT has manifest power counting in the typical velocity $v$ of the sources, and governs the dynamics of the long wavelength ``radiation" gravitons that propagate out towards the gravitational wave detector.  To illustrate our formalism, in this section we also give two classic examples of how the matching procedure and the EFT velocity power counting rules are applied to reproduce the leading non-trivial consequences of general relativity in the context of the NR two-body problem:   the Lagrangian that describes the gravitational interactions of two NR objects at ${\cal O}(v^2)$ beyond Newtonian gravity~\cite{EIH}, and the interactions of gravitational radiation with the binary star at leading order in the velocity.  

In Sec.~\ref{sec:RG} we return to the relativistic point particle effective Lagrangian, and show how all point particle divergences can be absorbed into worldline operators.  In Sec.~\ref{sec:matching}, we discuss how, given a more complete stellar model one can determine the coefficients of operators in the point particle EFT at the scale $\mu\sim 1/r_s$.  Finally, in Sec.~\ref{sec:v6log} we show how the RG running of the EFT coefficients at order $v^6$ can be exploited to yield a simplified calculation of the terms in the NR expansion of the two-body gravitational Lagrangian that scale as $v^6\ln\alpha$. This calculation serves as a prototype for the RG running of worldline operators.  We have also included
an appendix which illustrates how the Feynman rules of the EFT are derived and used.

\section{The Effective Theory}
\label{sec:EFT}

\subsection{Kinematics}

To construct an EFT that has manifest velocity power counting, we must first carefully analyze the kinematic configurations that arise in the binary problem.  Consider a binary  system composed of slowly moving neutron stars or black holes.  Let $m$ characterize the typical mass of the constituents (taking them, in this discussion, to be roughly the same size), $r$ the typical separation between them and $v$ their relative three-velocity.  In a system whose evolution is determined to leading order by a $1/r$ Newtonian interaction, these parameters are not independent, since the virial theorem relates 
\begin{equation}
 v^2 \sim {G_N m\over r}.
\end{equation}
The EFT for the binary system will contain degrees of freedom representing the point particle binary constituents coupled to gravitons.  While the typical particle momenta are of order $(E\sim mv^2, {\bf p}\sim mv)$, the gravitons appearing in a generic Feynman diagram have momenta that can be divided into two classes.  Gravitons with momentum scaling as $(k^0\sim v/r,{\bf k}\sim 1/r),$ mediate the forces responsible for binding the 2-body system.  These potential gravitons can never go on shell and thus will not appear as propagating degrees of freedom.  On the other hand, radiation gravitons with momentum $(k^0\sim v/r,{\bf k}\sim v/r)$ can appear on shell and must be kept in the EFT to reproduce the correct long distance physics.  

Note that the interaction of an NR particle with a single potential or radiation graviton causes the particle to recoil by a fractional amount $|{\bf k}|/|{\bf p}|\sim \hbar/L\ll 1$ , where $L\sim m v r$ is the typical orbital angular momentum of the system.  Thus as far as the graviton dynamics is concerned, the NR particles can be treated as background non-dynamical sources.  Consequently,  in this regard the EFT that describes the binary configuration has more in common with heavy particle effective field theories (eg., HQET~\cite{HQET}), in which light modes interact with a static, non-propagating source than a theory such as NRQCD~\cite{NREFT,LMR}, where the quark and gluon degrees of freedom have momentum components which differ only by powers of the velocity.  However, unlike HQET our EFT is
not an expansion in inverse powers of the mass.

As we will see, treating the stellar objects as non-propagating sources of gravitons also resolves a long standing formal problem in the treatment of general relativity as an EFT~\cite{GREFT,GREFTrev}.  For astrophysical systems with masses $m\gg m_{Pl}$, one would naively conclude from the conventional Feynman rules of the EFT that the expansion parameter is $m/m_{Pl}\gg 1$ and therefore the energy expansion implicit in the interpretation of gravity as an EFT seems to be spoiled~\cite{GRNRPC}.  In fact this obstruction to the EFT interpretation of general relativity is only an artifact of treating the point sources as dynamical degrees of freedom, and disappears when the theory is formulated in terms of worldlines interacting with gravitons.  Once this is done, the power counting in $v$ becomes manifest at the level of the  action.

\subsection{Point particle Effective Theory}

Given these remarks, the starting point of our EFT formulation consists of a theory of relativistic point particles coupled to gravity
\begin{equation}
\label{eq:ZOSO}
S= S_{EH} + S_{pp},
\end{equation}
where the Einstein-Hilbert term
\begin{equation}
S_{EH} = - 2 m_{Pl}^2 \int d^4x \sqrt{g} R(x),
\end{equation}
describes the graviton dynamics\footnote{Our conventions are $R_{\mu\nu} = \partial_\alpha{\Gamma^\alpha_{\mu\nu}} - \partial_\nu {\Gamma^\alpha_{\alpha\mu}} + \cdots$ and signature $(+,-,-,-)$.  $m_{Pl}^2=1/32\pi G_N$}  and
\begin{eqnarray}
\label{eq:pp}
\nonumber
S_{pp} &=& - \sum_a m_a \int d\tau_a + \sum_a c^{(a)}_R \int d\tau_a R(x_a)\\
 &+& \sum_a c^{(a)}_V \int d\tau_a R_{\mu\nu}(x_a) {\dot x}_a^\mu  {\dot x}_a^\nu + \cdots
\end{eqnarray}
determines the motion of the 2-body system ($a=1,2$ labels the different particles).  In this equation, $d\tau_a=\sqrt{g_{\mu\nu}dx_a^\mu dx^\nu_a}$ is the proper time along the worldline $x_a^\mu$ of the $a\mbox{-th}$ particle.  We will ignore in this paper additional degrees of freedom that describe the spin or any additional multipole moments carried by each particle.  This is justified for an extended object in which the energy cost of exciting internal degrees of freedom is large compared to the typical frequency of the gravitons which it interacts with.  Thus strictly speaking the formalism presented here can be used only to describe the dynamics of spinless black holes or spherically symmetric compact objects.   We will save the problem of including multipole moments in the EFT for future work.

The first term in $S_{pp}$ generates geodesic motion about the metric $g_{\mu\nu}$.   Besides this term, we have also explicitly shown the first two of an infinite set of possible non-minimal couplings of the point objects to the spacetime metric.  Since these non-minimal couplings cause deviations from pure geodesic motion, one would expect them to be associated with the effects of finite spatial extent of the sources.  Thus by including the most general set of such operators, we systematically take into account all possible corrections due to the finite size $r_s$ of the object.  As we discuss later, the operators linear in the curvature appearing in Eq.~(\ref{eq:pp}) will not contribute to any physical observable\footnote{  These operators have also been discussed in Ref.~\cite{de}}.  This should be contrasted with higher dimension operators in the point particle action involving higher powers of the curvature tensor, which encode tidal deformations of the binary constituents.  We will discuss in a Sec.~\ref{sec:matching} how one goes about determining the precise relation between the coefficients of these operators and the microscopic physics (ie, the stellar structure model) that determines the internal structure.  A discussion of the laws of motion for extended objects in general relativity can be found in~\cite{hartlethorne}.

A gravitational wave detector such as LIGO measures indirectly the power emitted in gravitational radiation from the binary system (see Eq.~(\ref{eq:phase})).  Given the action Eq.~(\ref{eq:ZOSO}), it is in principle possible to calculate this quantity by expanding the metric around flat space, $g_{\mu\nu} = \eta_{\mu\nu} + h_{\mu\nu}$, and integrating out the graviton field $h_{\mu\nu}$ to obtain an effective action for the particle coordinates alone
\begin{equation}
\label{eq:ZOSOPI}
\exp[i S_{eff}[x_a]] = \int{\cal D}h_{\mu\nu} \exp[i S_{EH} + iS_{pp}].
\end{equation}
The effective action $S_{eff}[x_a]$ has a real part which generates the coupled equations of motion for the 2-body system, and consequently its mechanical binding energy.  It also has an imaginary part that measures the total number of gravitons emitted by a fixed two-particle configuration $\{x^\mu_a\}$ over an arbitarily large time $T\rightarrow\infty$ 
\begin{equation}
{1\over T} \,\mbox{Im} S_{eff}[x_a] =  {1\over 2} \int dE d\Omega {d^2\Gamma\over dE d\Omega},
\end{equation}
where $d\Gamma$ is the differential rate for graviton emission from the binary system.  From this quantity we obtain the classical power spectrum $dP = E d\Gamma$, and therefore the gravitational wave phase seen by the detector.  

In principle, one could directly evaluate Eq.~(\ref{eq:ZOSOPI}) using the Lorentz covariant Feynman rules generated by the Einstein-Hilbert Lagrangian (the Feynman diagram expansion for perturbative gravity is reviewed, for example, in~\cite{GREFTrev,veltman}), taking the NR limit at the level of each amplitude.  However, the perturbative series generated in this way is not optimal for taking the limit $v\ll 1$.  For example, the one-graviton exchange term in $iS_{eff}[x_a]$,
\begin{equation}
\sum_{a\neq b} {m_a m_b\over 8 m_{Pl}^2} \int d\tau_a d\tau_b \left[1-2 ({\dot x}_a\cdot {\dot x}_b)^2\right] D_F(x_a-x_b)
\end{equation}   
contains an infinite number of terms from the NR expansion of the proper times $d\tau^2_a = \eta_{\mu\nu} dx^\mu_a dx^\nu_a$ and the Feynman propagator $D_F(x)$.  While for a term like this it is easy to do the expansion in small velocity, in a more complicated diagram with multiple internal graviton lines it becomes more cumbersome to keep track of all the necessary terms at a given order in $v$.  Furthermore suppose one is interested, because of the limited sensitivity of the detector, in computing an observable only to some fixed order in $v$.  Then how many diagrams do we need to keep?  It is not possible to answer this question with the Feynman rules generated by Eq.~(\ref{eq:ZOSO}) as it stands.  Likewise, given the non-minimal tidal couplings in Eq.~(\ref{eq:pp}) it is not possible to determine at what order in $v$ they may first contribute.  Since from the point of view of the point particle theory, Eq.~(\ref{eq:ZOSO}), the NR dynamics occurs in the far infrared, to formulate an EFT that is tailored to the limit $v\ll 1$, we must integrate out all field degrees of freedom with wavelengths shorter than the orbital distance scale $r$.  We now turn to a formulation of this EFT.     

\subsection{NRGR}

The EFT that systematically describes the NR two-body problem, which we call NRGR, has manifest power counting in the velocity $v$.  To construct NRGR, we proceed in several steps.  We work in a preferred set of frames in which the relative velocity is small, allowing us to expand in $v$.  Then in order to have interaction vertices with homogeneous  velocity scaling, we simply expand $S_{pp}$ in powers of the particle three-velocities.  Ignoring for now the effects of the higher-dimension tidal operators, this leads to a worldline Lagrangian
\begin{eqnarray}
\nonumber
L_{pp} &=& \sum_a m_a \left[ {1\over2} {\bf v}^2_a -{1\over 2} h_{00} -  h_{0i} {{\bf v}_a}_i  - {1\over 4} h_{00} {\bf v}^2_a\right.\\
& & \left. {} - {1\over 2} h_{ij} {{\bf v}_a}_i {{\bf v}_a}_i  + {1\over 8}  {\bf v}^4_a  +\cdots\right],
\end{eqnarray}
where $h_{00},h_{0i},h_{ij}$ are evaluated on the point particle worldline $(x^0,{\bf x}_a(x^0))$ (we use the Euclidean metric $\delta_{ij}$ to raise/lower spatial indices $i,j=1,2,3$).  

The propagator for the field $h_{\mu\nu}$ appearing here is still fully relativistic, so it does not distinguish between potential gravitons, which have no right to appear as degrees of freedom in NRGR, and radiation gravitons.  Because of this, the Feynman rules still do not scale homogeneously with $v$.  To deal with this problem, it is convenient to decompose the graviton as
\begin{equation}
h_{\mu\nu}(x) = {\bar h}_{\mu\nu}(x) + H_{\mu\nu}(x),
\end{equation} 
where $H_{\mu\nu}$ represents the potential gravitons, with 
\begin{eqnarray}
\partial_i H_{\mu\nu}\sim {1\over r} H_{\mu\nu} & &  \partial_0 H_{\mu\nu}\sim {v\over r} H_{\mu\nu} ,
\end{eqnarray}
and ${\bar h}_{\mu\nu}$ describes a long-wavelength radiation field
\begin{equation}
\partial_\alpha {\bar h}_{\mu\nu} \sim {v\over r} {\bar h}_{\mu\nu}.
\end{equation}
Actually, it is better to further decompose $H_{\mu\nu}$ by removing from it the large momentum fluctuations ${\bf k}\sim 1/r$.  To do this we simply work with the Fourier transformed field \cite{LMR} $H_{{\bf k}\mu\nu}(x^0)$ 
\begin{equation}
H_{\mu\nu}(x) = \int_{\bf k} e^{i{\bf k}\cdot {\bf x}} H_{{\bf k}\mu\nu}(x^0),
\end{equation}
where $\int_{\bf k}$ is shorthand for $\int {d^3{\bf k}/(2\pi)^3}$.  The advantage of this redefinition is that now derivatives acting on any field in the EFT scale in the same way, $\partial_\mu\sim v/r$, so it is easy to count powers of $v$ coming from derivative interactions.

The NRGR lagrangian can then be derived by computing the functional integral
\begin{equation}
\label{eq:NRGRPI}
\exp[i S_{NRGR}[x_a,{\bar h}]] = \int {\cal D}H_{\mu\nu} \exp[i S[{\bar h}+H,x_a] + i S_{GF}],
\end{equation}
where $S_{GF}$ is a suitable gauge fixing term.  We have not included ghost terms because as we will see, they are not needed in any NRGR computation.  Eq.~(\ref{eq:NRGRPI}) indicates that as far as the potential modes $H_{\mu\nu}$ are concerned ${\bar h}_{\mu\nu}$ is just a slowly varying background field.  To preserve gauge invariance of the NRGR action, we will chose $S_{GF}$ to be invariant under general coordinate transformations of the background metric  ${\bar g}_{\mu\nu}(x) = \eta_{\mu\nu} + {\bar h}_{\mu\nu}(x)$.  To be definite, we take the gauge
\begin{equation}
\label{eq:pgauge}
S_{GF} = m^2_{Pl}\int d^4 x \sqrt{\bar g} \Gamma_\mu \Gamma^\mu,
\end{equation}
with $\Gamma_\mu = D_\alpha H^\alpha_\mu - {1\over 2}D_\mu H^\alpha_\alpha$, where $D_\mu$ is the covariant derivative derived from the background metric ${\bar g}_{\mu\nu}$.  In this gauge, the ${\cal O}(H^2)$ terms in the action are  (after performing a rescaling of $H_{\mu\nu}$ to obtain a canonically normalized kinetic term)
\begin{eqnarray}
\nonumber
{\cal L}_{H^2} &=& -{1\over 2} \int_{\bf k}\left[{\bf k}^2 H_{{\bf k}\mu\nu} H_{-{\bf k}}^{\mu\nu}-{{\bf k}^2\over 2} H_{\bf k}  H_{-{\bf k}}\right.\\
& & \left. -\partial_0 H_{{\bf k}\mu\nu} \partial_0 H_{-{\bf k}}^{\mu\nu} + {1\over 2} \partial_0 H_{\bf k}  \partial_0 H_{-\bf k}\right],
\end{eqnarray}
where $H_{\bf k}=H^\mu_{\mu \bf k}$.  The terms in the second line of this equation are suppressed relative to the first line by a power of $v^2$, and are treated perturbatively, as operator insertions, in correlation functions.  Thus, the propagator for $H_{{\bf k}\mu\nu}$ is 
\begin{equation}
\label{eq:potprop}
\langle H_{{\bf k}\mu\nu} (x^0)H_{{\bf q}\alpha\beta}(0)\rangle = -(2\pi)^3\delta^3({\bf k} + {\bf q}){i\over {\bf k}^2}\delta(x_0)P_{\mu\nu;\alpha\beta}, 
\end{equation}
where $P_{\mu\nu;\alpha\beta} = {1\over 2}\left[\eta_{\mu\alpha} \eta_{\nu\beta} + \eta_{\mu\beta} \eta_{\nu\alpha} - {2\over d-2}\eta_{\mu\nu} \eta_{\alpha\beta}\right]$ with $d$ the spacetime dimension.  If we assign the scaling $x^0\sim r/v$, ${\bf k}~\sim 1/r$, we learn from this that a potential graviton scales as $H_{{\bf k}\mu\nu}\sim v^{1/2} r^2$.  Likewise, because in mometum space the radiation graviton propagator scales as $1/k^2\sim r^2/v^2$, the coordinate space radiation graviton field is taken to scale as ${\bar h}_{\mu\nu}\sim v/r$.

\begin{figure}
    \centering
    \includegraphics[width=4cm]{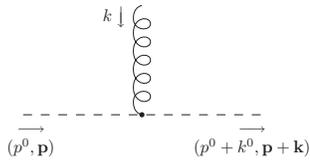}
\caption{The incoming potential graviton with momentum ${\bf p}$ absorbs a radiation graviton with
momentum $k$.}
\label{fig:mult}
\end{figure}

Just as we had to expand $H_{\mu\nu}$ in terms of momentum labels to obtain manifest power counting~\cite{LMR}, it is also necessary to perform a multipole expansion of the radiation field ${\bar h}_{\mu\nu}$ whenever it couples to particle worldlines or to potential modes~\cite{QCDmult}.  To see the necessity for this, consider the absorbtion of a radiation graviton from a potential graviton as shown in Fig.~\ref{fig:mult}. The outgoing potential graviton has a propagator of the form
\begin{equation}
{1\over ({\bf p}+{\bf k})^2}
\end{equation}
which does not scale homogenously in $v$ since ${\bf p}\sim {\cal O}(1/r)$ and ${\bf k} \sim {\cal O}(v/r)$. The propagator must be expanded in powers of $|{\bf k}|/|{\bf p}|$, which is achieved by multipole expanding the interactions of the radiation graviton field with the potentials at the level of the action,
\begin{eqnarray}
\label{eq:mult}
\nonumber
{\bar h}_{\mu\nu}(x^0,{\bf x}) &=& {\bar h}_{\mu\nu}(x^0,{\bf X}) + \delta {\bf x}_i \partial_i {\bar h}_{\mu\nu}(x^0,{\bf X}) \\
 & & {} + {1\over 2}\delta {\bf x}_i \delta {\bf x}_j \partial_i\partial_j {\bar h}_{\mu\nu}(x^0,{\bf X}) + \cdots,
\end{eqnarray}
where $\delta {\bf x} = {\bf x} -{\bf X}$, and ${\bf X}$ is some reference point in the vicinity of the point particle ensemble, for instance the center of mass.  Likewise, consider the amplitude for the point particle system to absorb a $00$-polarized radiation graviton with 4-momentum $k$
\begin{equation}
i{\cal A} =  -{i\over 2}\sum_a {m_a\over m_{Pl}}\int dx^0 \epsilon_{00}(k) e^{-i k\cdot x_a}.
\end{equation}
A graviton emitted during the binary inspiral phase has 4-momentum $k\sim (v/r,v/r)$, so if the coordinates ${\bf x}_a$ are measured relative to, for instance, the center of mass, then we have ${\bf x}_a/x^0\sim v$ and consequently this amplitude contains an infinite number of terms each with a different velocity scaling.  This situation is also remedied by the multipole expansion in Eq.~(\ref{eq:mult}).  
\begin{table}
\begin{eqnarray}
\nonumber
\begin{array}{c|c|c|c}
  {\bf k} & H^{\bf k}_{\mu\nu} & {\bar h}_{\mu\nu}  & m/m_{Pl}\\
\hline
   1/r      & r^2 v^{1/2}                   & v/r & \sqrt{L v}\\
\end{array}
\end{eqnarray}
\caption{NRGR power counting rules.}
\label{tab}
\end{table}

\begin{figure*}[t!]
    \centering
    \includegraphics[width=0.9\textwidth]{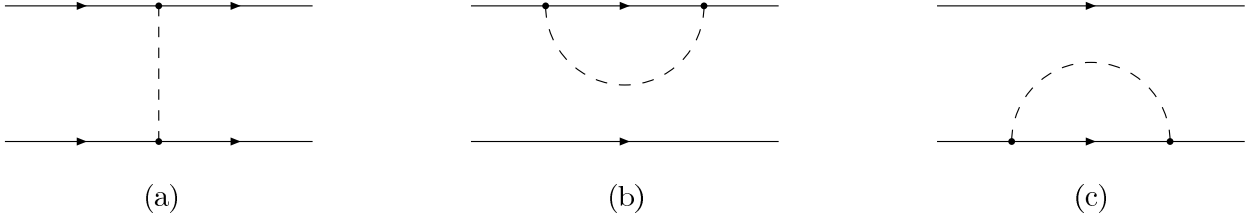}
\caption{Diagrams contributing to $Lv^0$ potentials.  The self-energy graphs in (b), (c) are pure counterterm and have no physical effect.}
\label{fig:newt}
\end{figure*}

These field redefinitions are enough to recast the original point particle theory of Eq.~(\ref{eq:ZOSO}) into a form that is better suited to a small velocity expansion.  The power counting rules in Table~\ref{tab} allow one to determine the order in the velocity at which any diagram in the calculation of the functional integral in Eq.~(\ref{eq:NRGRPI}), as well as in Feynman diagrams computed in NRGR.  However the simplest example of the power counting rules, the counting of the particle NR kinetic term, reveals a slight subtlety that we must deal with.  Given that $dx^0\sim r/v$, we have
\begin{equation}
\int dx^0 m {\bf v}^2 \sim {r\over v} m v^2 = m v r.
\end{equation}  
Thus this term scales as the orbital angular momentum $L\gg 1$.  This is not the only term that scales in this way.  The exchange of a single $H_{00}$ potential graviton between two particles, responsible for the Newton force, involves the computation of the diagram of Fig.~\ref{fig:newt}a, which is given by
\begin{eqnarray}
\label{eq:newta}
\nonumber
\mbox{Fig.~\ref{fig:newt}a} &=&  {i m_1 m_2 \over 8 m^2_{Pl}}\int dt_1 dt_2  \delta(t_1-t_2) \int_{\bf k} {1\over {\bf k}^2} e^{-i{\bf k}\cdot\left({\bf x}_1 - {\bf x}_2\right)} \\
 &=& i \int dt {G_N m_1 m_2\over |{\bf x}_1(t) - {\bf x}_2(t)|},
\end{eqnarray}  
($G_N\equiv 1/32\pi m^2_{Pl}$).  This scales as  
\begin{equation}
[(dt) ({d^3\bf k}) (m/m_{Pl}) H]^2\sim mvr, 
\end{equation}
by virtue of the virial theorem (the power divergent self-energy graphs Fig.~\ref{fig:newt}(b), Fig.~\ref{fig:newt}(c), which renormalize the particle masses, vanish when evaluated in dimensional regularization).  It can be shown in general that all other contributions to $S_{eff}[x_a]$ are down by powers of $v$ relative to the particle kinetic terms or the Newton potential (see the appendix).  It is then easy to deal with the presence of the large parameter $L$ in the power counting.  We must treat the operators that scale as $L v^0$ non-perturbatively, or in other words, we must expand about background fields ${\bf x}_a$ that to leading order in $v$ satisfy the  equations of motion of Newtonian gravity, treating operators which scale  as $L v^n$ perturbatively.  This is the resolution of the problem raised in~\cite{GRNRPC} regarding the apparent breakdown of a manifest perturbative expansion when the particle masses are larger than $m_{Pl}$.   

One can also show that diagrams that contain graviton loops are supressed by powers of $1/L$ relative to the tree diagrams (after particle worldlines are stripped off).  This is in accord with the intuition that loop diagrams represent quantum corrections which should not be kept in the perturbative expansion of classical field theory observables.  The fact that loop corrections are irrelevant to the quantities of interest also explains why we did not need to keep ghost terms in the functional integral of Eq.~(\ref{eq:NRGRPI}).

What is the resulting theory after all potential gravitons are integrated out?  It is a theory of radiation gravitons coupled to moments of the two-particle distribution in a gauge invariant manner.  Working in the CM frame, $\sum_a m_a {\bf x}_a=0$, the leading order contributions to processes in NRGR can be calculated from
\begin{eqnarray}
\label{eq:NRGRL}
L &=& {1\over 2} \sum_a m_a {\bf v}^2_a +  {G_N m_1 m_2\over |{\bf x}_1-{\bf x}_2|} -
{m \over 2 m_{Pl}} {\bar h}_{00} \\
\nonumber
& & {} -{{\bar h}_{00}\over 2 m_{Pl}} \left[{1\over 2} \sum_a m_a{\bf v}^2_a - {G_N m_1 m_2\over |{\bf x}_1 -{\bf x}_2|}\right]\\
\nonumber
& &  -{1\over 2 m_{Pl}} \epsilon_{ijk} {\bf L}_k\partial_j {\bar h}_{0i} + {1\over 2 m_{Pl}}\sum_a m_a {\bf x}_{ai}{\bf x}_{aj} R_{0i0j},
\end{eqnarray}
where we suppress the argument $(x^0,{\bf X}_{cm})$ of the canonically normalized radiation field ${\bar h}_{\mu\nu}$, and write $m=\sum_a m_a,$ ${\bf X}_{cm} = {1\over m}\sum_a m_a {\bf x}_a$, ${\bf L} = \sum_a m_a {\bf x}_a\times {\bf v}_a$.  Note that after integrating out the potential modes, the only dependence on the orbital scale $r$ appears in the Wilson coefficients (the moments) of the theory.  Thus in our theory, the decoupling between near-zone NR physics and far-zone radiation is manifest at the level of the NRGR Lagrangian.  This explicit separation of scales, which is what motivates the EFT approach, should simplify the evaluation of higher order radiative tail effects within our formalism.  Of the terms shown here, only the coupling of the (linearized) Riemann tensor to the moment $I_{ij}=\sum_a m_a {{\bf x}_a}_i {{\bf x}_a}_j,$ can source gravitational radiation, since all other terms in this Lagrangian couple to conserved quantities and therefore cannot radiate, up to higher order radiation reaction effects.  In fact, the trace of the linearized Riemann tensor
\begin{equation}
R_{i0i0} = R_{00} = \partial_0 \partial_\mu {\bar h}^\mu_0 -{1\over 2}\partial_\mu \partial^\mu {\bar h}_{00} -{1\over 2}\partial_0^2 {\bar h} + {\cal O}({\bar h}^2)
\end{equation}   
vanishes for on-shell graviton  matrix elements, so that radiation only couples to the traceless quadrupole moment of the source
\begin{equation}
{Q}_{ij} = \sum_a m_a \left({{\bf x}_a}_i {{\bf x}_a}_j -{1\over 3}\delta_{ij} {\bf x}^2_a\right).
\end{equation}
In the next section we will show more explicitly how to derive Eq.~(\ref{eq:NRGRL}).

\begin{figure}
    \centering
    \includegraphics[width=4cm]{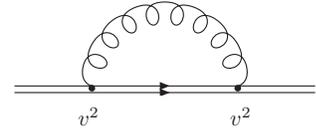}
\caption{NRGR diagram whose imaginary part gives rise to the quadrupolar gravitational radiation power spectrum.}
\label{fig:selfrad}
\end{figure}

Given Eq.~(\ref{eq:NRGRL}), how do we obtain the power radiated in gravitational waves from this formalism?  We simply calculate the imaginary part of the self-energy diagram of Fig.~\ref{fig:selfrad},
\begin{eqnarray}
\mbox{Fig.~\ref{fig:selfrad}} &=& -{1\over 8 m^2_{Pl}} \int dx^0_1 dx^0_2 I_{ij}(x^0_1) I_{kl}(x^0_2)\\
\nonumber
& & \times \langle R_{0i0j}(x^0_1,{\bf X}_{cm}) R_{0k0l}(x^0_2,{\bf X}_{cm})\rangle.
\end{eqnarray}
(We have only kept the part of the diagram that arises from the coupling $I_{ij} R_{0i0j}$, since it is the only term at this order which has a non-zero imaginary part).  Computing the two-point function of the linearized $R_{0i0j}$ one finds that it is proportional to the projection operator onto symmetric and traceless two-index spatial tensors, ${1\over 2}\delta_{ik}\delta_{jl} + {1\over 2}\delta_{il}\delta_{jk} -{1\over 3} \delta_{ij} \delta_{kl}$.  Writing the above equation in terms of the Fourier transform of the quadrupole moment
\begin{equation}
Q_{ij}(k_0) = \int dx^0 Q_{ij}(x^0) e^{-i k_0 x^0},
\end{equation} 
we end up with
\begin{equation}
\mbox{Fig.~\ref{fig:selfrad}}= -{i\over 80 m^2_{Pl}}\int {d^4 k\over (2\pi)^4} {k^4_0\over k^2 +i\epsilon} |Q_{ij}(k_0)|^2.
 \end{equation}
This diagram then gives rise to the following contribution to $\mbox{Im} S_{eff}[x_a]$
\begin{equation}
\mbox{Im} S_{eff}  = -{1\over  80 m^2_{Pl}}\int {d^3{\bf k}\over (2\pi)^3 2|{\bf k}|} {\bf k}^4 |Q_{ij}(|{\bf k}|)|^2.
\end{equation}
The integrand in this equation is proportional to the differential graviton emission rate over the history of the two-particle system.  To get the differential power, multiply the integrand by an additional factor of the graviton energy $|{\bf k}|$.  In particular, including this extra factor of the energy and integrating over all momenta, we obtain the total power radiated in gravitational waves  
\begin{eqnarray}
\nonumber
P &=&  {G_N\over 5\pi T} \int_0^\infty d\omega \omega^6 |Q_{ij}(\omega)|^2\\
   &=& {G_N\over 5}\langle \dddot{Q}_{ij} \dddot{Q}_{ij}\rangle,
\end{eqnarray}
where the dots denote time derivatives, and the brackets denote a time average.  Of course, this equation is just the usual quadrupole radiation formula.  

\begin{figure*}[t!]
    \centering
    \includegraphics[width=0.9\textwidth]{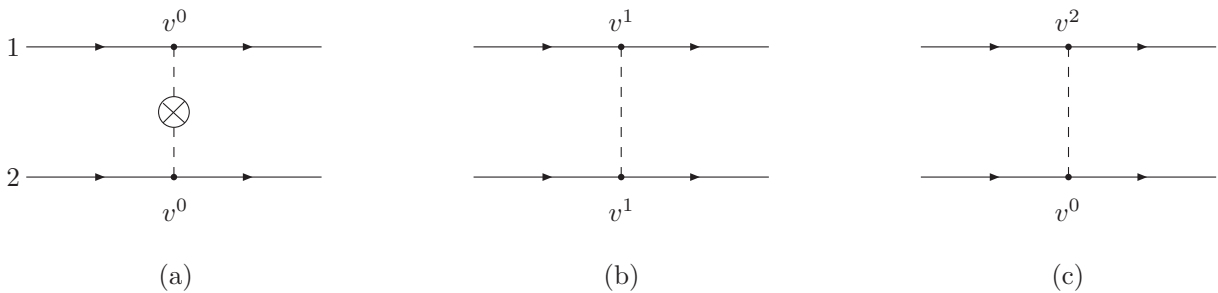}
\caption{Diagrams contributing to $Lv^2$ terms in the two-body potential.  The $\otimes$ in (a) denotes an insertion of the potential graviton kinetic term.  A similar diagram to (c) with $1\leftrightarrow 2$ is not shown.}
\label{fig:EIH1}
\end{figure*}

\subsection{Examples at NLO}

The construction of the EFT in the previous section was perhaps somewhat formal, so it is helpful to see examples of how to explicitly calculate the terms in the NRGR action.  We will work out two textbook examples in this section:  the Lagrangian that leads to the leading order predictions for the power radiated in gravitational waves, Eq.~(\ref{eq:NRGRL}), and the EIH Lagrangian~\cite{EIH} that describes the ${\cal O}(v^2)$ general relativity corrections to the gravitational interaction of two point masses.

\subsubsection{EIH Lagrangian}

First consider $L_{EIH}$.  It arises from the diagrams shown in Fig.~\ref{fig:EIH1} and Fig.~\ref{fig:EIH2}.  Using the NRGR Feynman rules, the sum of the diagrams with a single potential graviton exchange are given by
\begin{eqnarray}
\label{eq:EIH1}
\nonumber
\mbox{Fig.~\ref{fig:EIH1}a} &=& {i m_1 m_2\over 8 m^2_{Pl}}\int dt_1 dt_2 {d^2\over dt_1 dt_2}\delta(t_1-t_2)\\
\nonumber
& & {} \times \int_{\bf k} {1\over {\bf k}^4} e^{-i{\bf k}\cdot({\bf x}_1 -{\bf x}_2)}\\
\nonumber
&=&  {i\over 2}\int dt {G_N m_1 m_2\over  |{\bf x}_1 - {\bf x}_2|}\\
& & \times \left[{\bf v}_1\cdot {\bf v}_2 - {({\bf v}_1\cdot {\bf x}_{12} )({\bf v}_2\cdot {\bf x}_{12})\over |{\bf x}_1 -{\bf x}_2|^2} \right],
\end{eqnarray}
with ${\bf x}_{12}={\bf x}_1-{\bf x}_2$,
\begin{eqnarray}
\label{eq:EIH2}
\nonumber
\mbox{Fig.~\ref{fig:EIH1}b} &=& {im_1 m_2\over  m^2_{Pl}}\int dt_1 dt_2 \delta(t_1-t_2) {{\bf v}_1}_i {{\bf v}_2}_j \\
\nonumber
& & {} \times \int_{\bf k} {1\over {\bf k}^2} P_{0i;0j} e^{-i{\bf k}\cdot({\bf x}_1 -{\bf x}_2)} \\
&=& - 4 i\int dt {G_N m_1 m_2\over |{\bf x}_1 -{\bf x}_2|}({\bf v}_1\cdot {\bf v}_2),
\end{eqnarray}
and
\begin{eqnarray}
\label{eq:EIH3}
\nonumber
\mbox{Fig.~\ref{fig:EIH1}c} &=& {im_1 m_2\over 4 m^2_{Pl}}\int dt_1 dt_2 \delta(t_1-t_2)\left[{1\over 4} {\bf v}^2_1\right.\\
\nonumber & & {} \left. + {{\bf v}_1}_i {{\bf v}_1}_j P_{ij;00}  \right]  \int_{\bf k} {1\over {\bf k}^2}  e^{-i{\bf k}\cdot({\bf x}_1 -{\bf x}_2)} \\
&=& {3i\over 2}\int dt {G_N m_1 m_2\over |{\bf x}_1 -{\bf x}_2|} {\bf v}^2_1.
\end{eqnarray}
Here, we have used the formula
\begin{equation}
\int {d^d{\bf k}\over (2\pi)^d} {1\over ({\bf k}^2)^\alpha} e^{-i{\bf k}\cdot {\bf x}} = {1\over (4 \pi)^{d/2}} {\Gamma(d/2-\alpha)\over \Gamma(\alpha)} \left({{\bf x}^2\over 4}\right)^{\alpha-d/2}.
\end{equation}

However, at this order in $v$, the single exchange diagrams are not sufficient.  The NRGR power counting tells us that we also get contributions from diagrams with two and three graviton propagators.  For example, the graph in Fig.~\ref{fig:EIH2}a scales as 
\begin{widetext}
\begin{equation}
\mbox{Fig.~\ref{fig:EIH2}a} \sim \left[dx^0 d^3{\bf k}\left({m\over m_{Pl}}\right) H_{00}\right]^3 \left[dx^0 (d^3{\bf k})^3 \delta^3({\bf k}) {{\bf k}^2 H^3_{00}\over m_{Pl}}\right] \sim \left[{r\over v}{1\over r^3} (L v)^{1/2} r^2 v^{1/2}\right]^3\left[{1\over m_{Pl}}{r\over v} {1\over r^8} \left(r^2 v^{1/2}\right)^3\right]  \sim L v^2.
\end{equation}
\end{widetext}
The first factor stems from three insertions of leading order vertices while the second bracket arises from an insertion of a three-graviton interaction.  Expanding the Einstein-Hilbert Lagrangian to order $H^3$, it is straightforward to show
\begin{widetext}
\begin{equation}
\label{eq:3grav}
\langle T( H_{{\bf k}_1}^{00}(x_1)  H_{{\bf k}_2}^{00}(x_2)  H_{{\bf k}_3}^{00}(x_3))\rangle = - {i\over 4 m_{Pl}}\delta(x^0_1-x^0_2)\delta(x^0_1-x^0_3) (2\pi)^3\delta^3\left(\sum_r {\bf k}_r\right) \prod_{r=1}^3 {i\over {\bf k}_r^2}\times \sum_{r=1}^3 {\bf k}^2_r,
\end{equation}
so that
\begin{equation}
\label{eq:EIH4}
\nonumber
\mbox{Fig.~\ref{fig:EIH2}a} = {i m_1 m^2_2\over 16 m^3_{Pl}}\int dx^0_1 dx^0_2 dx^0_{2'} \int_{{\bf k}_1, {\bf k}_2, {\bf k}_3} e^{i\sum_i {\bf k}_i \cdot {\bf x}_i} \langle  T(H_{{\bf k}_1}^{00}(x_1)  H_{{\bf k}_2}^{00}(x_2)  H_{{\bf k}_3}^{00}(x_{2'}))\rangle = -i\int dt {G^2_N m_1 m^2_2\over |{\bf x}_1(t) -{\bf x}_2(t) |^2},
\end{equation} 
where we have dropped linearly UV divergent self-energy contributions which can be absorbed into the particle masses, and which vanish when evaluated in dimensional regularization.  Finally, the seagull diagram, Fig.~\ref{fig:EIH2}b, also contributes at order $Lv^2$, and is given by
\begin{figure*}[t!]
    \centering
    \includegraphics[width=0.5\textwidth]{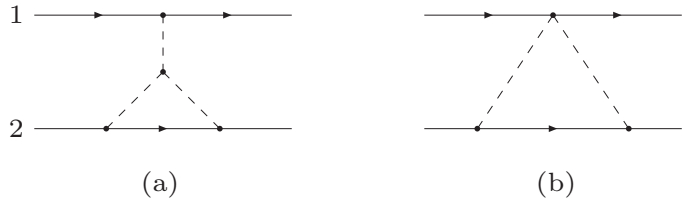}
\caption{Diagrams with two and three potential graviton lines that also contribute to the $Lv^2$potentials.  Similar diagrams with $1\leftrightarrow 2$ not shown.}
\label{fig:EIH2}
\end{figure*} 
\begin{equation}
\label{eq:EIH5}
\mbox{Fig.~\ref{fig:EIH2}b} = {i m_1 m^2_2\over 128 m^4_{Pl}}\int dt_1 dt_2 dt_{2'} \delta(t_1-t_2)\delta(t_1 - t_{2'})\left[\int_{{\bf k}} {1\over {\bf k}^2}e^{-i{\bf k}\cdot ({\bf x}_1 - {\bf x}_2)}\right]^2 = {i\over 2}\int dt {G^2_N m_1 m^2_2\over |{\bf x}_1(t) -{\bf x}_2(t)|^2}
\end{equation}
Putting together Eq.~(\ref{eq:EIH1}), Eq.~(\ref{eq:EIH2}), Eq.~(\ref{eq:EIH3}), Eq.~(\ref{eq:EIH4}), and  Eq.~(\ref{eq:EIH5}) together with their mirror images under the interchange of the particle labels, we end up with 
\begin{eqnarray}
L_{EIH} = {1\over 8}\sum_a m_a {\bf v}^4_a + {G_N m_1 m_2\over 2 |{\bf x}_1 -{\bf x}_2|}\left[3 ({\bf v}^2_1+ {\bf v}^2_2) - 7({\bf v}_1\cdot {\bf v}_2)  - {({\bf v}_1\cdot  {\bf x}_{12} )({\bf v}_2\cdot {\bf x}_{12})\over |{\bf x}_1-{\bf x}_2|^2} \right]- {G^2_N m_1 m_2 (m_1+m_2)\over 2  |{\bf x}_1 -{\bf x}_2|^2} ,
\end{eqnarray}
\end{widetext}
where we have also included the relativistic corrections to the kinetic energy of the point particles, which contribute to the effective action at order $Lv^2$.  This Lagrangian was first derived by Einstein, Infeld and Hoffmann~\cite{EIH}, by different methods.  A diagrammatic derivation of the static $1/r^2$ part of this Lagrangian, using Lorentz covariant worldline methods~\cite{worldline}, was given in ref.~\cite{muzinich}.  A quantum field theoretic derivation of the $1/r^2$ potential, treating the source particles as dynamical fields, was presented in~\cite{GREFT,EIHquant}.

\subsubsection{Couplings of Radiation Gravitons}

It is also instructive to use the formalism to derive the leading order Lagrangian that describes the interactions of radiation gravitons with NR sources, Eq~(\ref{eq:NRGRL}).  The derivation involves all aspects of the construction of NRGR (the multipole expansion for radiation, integrating out potential gravitons, velocity power counting rules).  At the lowest two orders in the expansion, the interaction of radiation with the NR sources follows simply from the multipole expansion of the radiation field. 
Expanding the proper time time term in powers of $v$, using the counting rules in Table I, leads to

\begin{eqnarray}
L_{v^{1/2}}  &=&-{1\over 2 m_{Pl}}\sum_a m_a{\bar h}_{00}.\\  
L_{v^{3/2}} &=&  - {1\over m_{Pl}} {{\bf P}_{cm}}_i {\bar h}_{0i}=0,
\end{eqnarray}
 in the CM frame.
 Note that to preserve manifest scaling  in $v$ we must multipole expand about the center of mass, as defined below Eq.~(\ref{eq:NRGRL}).  These terms (when integrated over time to form an action) scale as $\sqrt{L} v^{1/2}$, and $\sqrt{L} v^{3/2}$ respectively according to the power counting rules of the EFT, and generate diagrams which (i.e. terms in $S_{eff}[x_a]$) that are proportional to a single power of $L$. 

\begin{figure*}[t!]
    \centering
    \includegraphics[width=0.9\textwidth]{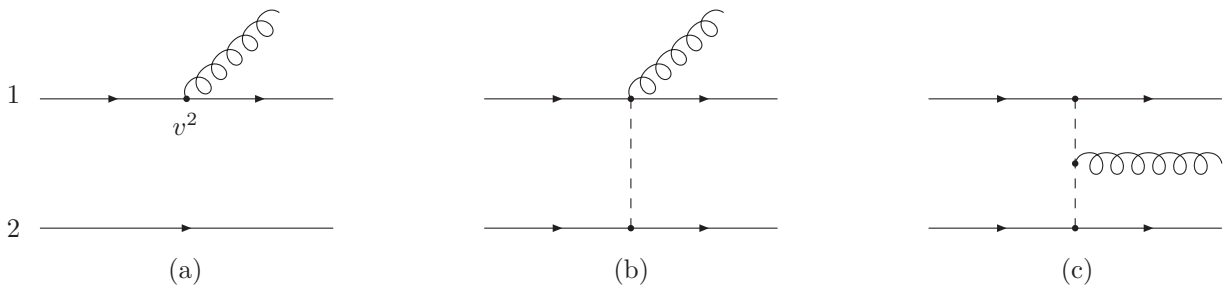}
\caption{Diagrams that contribute to the matching of radiation graviton couplings.  The diagram (a) is the multipole expansion carried out to order $v^2$.  Diagrams (b), and (c) are needed to ensure gauge invariance of the radiation graviton couplings.  Similar diagrams with $1\leftrightarrow 2$ not shown.}
\label{fig:jeebus}
\end{figure*}

At the next order in the velocity, the situation is not as straightforward.  Keeping terms with one more power of $v$ (either from the multipole expansion or from explicit powers of the particle velocities) one finds, in the CM frame,
\begin{eqnarray}
\nonumber
\label{eq:jeebus1}
L_{mult} &=&-\sum_a {m_a\over 2 m_{Pl}}\left[{1\over 2} {\bar h}_{00} {\bf v}^2_a+ {1\over 2}{{\bf x}_a}_i {{\bf x}_a}_j \partial_i\partial_j {\bar h}_{00}\right.\\
& & \left.{} + 2 {{\bf x}_a}_i {{\bf v}_a}_j \partial_i {\bar h}_{0j} + {\bar h}_{ij} {{\bf v}_a}_i {{\bf v}_a}_j\right].
\end{eqnarray} 
This Lagrangian has two problems.  First, it is not gauge invariant under infinitesimal coordinate transformations.   Second, it seems to predict that unphysical $00,0i$ graviton polarizations can be emitted by the two-particle bound state.  Of course, these two problems are related.  The resolution to this is that at this order in the expansion there are additional contributions to the effective Lagrangian, from diagrams with both external radiation gravitons and internal potentials, such as those of Fig.~\ref{fig:jeebus}.  The diagram in Fig.~\ref{fig:jeebus}b is  given by
\begin{equation}
\label{eq:jeebus2}
\mbox{Fig.~\ref{fig:jeebus}b} =  -{i\over 2}\int dx^0 {G_N m_1 m_2\over |{\bf x}_1 -{\bf x}_2|} {\bar h}_{00}.
\end{equation}
For the diagram in Fig.~\ref{fig:jeebus}c, we need the ${\bar h} H^2$, vertex.  It can be obtained from
\begin{widetext}
\begin{eqnarray}
{\cal L}_{{\bar h} H^2} &=&  {\bf k}^2 {\bar h}_{00} \left[-{1\over 4} H_{\bf k}^{\mu\nu} {H_{-{\bf k}}}_{\mu\nu} + {1\over 8} H_{\bf k}  H_{-{\bf k}} + {H_{\bf k}}^{\mu 0} {H_{-{\bf k}}}_{\mu 0} -  {1\over 2} {H_{\bf k}}_{00} H_{\bf -k}\right] + {\bf k}^2 {\bar h}_{0i}\left[2  H^{00}_{\bf k} {H_{-\bf k}}_{0i} - {H_{\bf k }}_{0i} H_{-\bf k}\right]\\
\nonumber
& & + {\bar h}_{ij} \left[{1\over 2} {\bf k}_i {\bf k}_j  H_{\bf k}^{\mu\nu} {H_{-{\bf k}}}_{\mu\nu} + {\bf k}^2 {H_{\bf k}}_{i \mu} {H_{-{\bf k}}}_j^\mu - {1\over 2}{\bf k}^2  {H_{\bf k}}_{ij} H_{\bf -k} - {1\over 4} {\bf k}_i {\bf k}_j  H_{\bf k}  H_{-{\bf k}} -\delta_{ij}{\bf k}^2 \left(-{1\over 4} H_{\bf k}^{\mu\nu} {H_{-{\bf k}}}_{\mu\nu} + {1\over 8} H_{\bf k}  H_{-{\bf k}}\right) \right],
\end{eqnarray}
\end{widetext}
where the integral over momentum ${\bf k}$ has been suppressed, and because of the multipole expansion, ${\bar h}_{\mu\nu}$ is evaluated at $(x^0,{\bf X}_{cm})$.  Given this term, it is easy to show that
\begin{eqnarray}
\nonumber
\mbox{Fig.~\ref{fig:jeebus}c} &=& {i m_1 m_2\over 8 m^3_{Pl}}\int dx^0\int_{{\bf k}} e^{-i{\bf k}\cdot ({\bf x}_1-{\bf x}_2)} {1\over {\bf k}^4} \\
& & \times  \left[{3\over 2} {\bf k}^2 {\bar h}_{00} +{1\over 2}{\bf k}^2  {\bar h}_{ii} - {\bf k}_i {\bf k}_j {\bar h}_{ij}\right].
\end{eqnarray}  
Using 
\begin{equation}
\int_{\bf k} e^{-i{\bf k}\cdot {\bf x}} {{\bf k}_i {\bf k}_j \over {\bf k}^4}  = {1\over 8\pi |{\bf x}|} \left[\delta_{ij} - {{\bf x}_i {\bf x}_j\over |{\bf x}|^2}\right], 
\end{equation}
as well as the equations of motion for ${\bf x}_{1,2}(t)$ from the order $Lv^0$ effective action, we find
\begin{eqnarray}
\label{eq:jeebus3}
\nonumber
\mbox {Fig.~\ref{fig:jeebus}c} &=& {i\over m_{Pl}} \int dx^0 \left[{3 G_N m_1 m_2\over 2 |{\bf x}_1 -{\bf x}_2|} {\bar h}_{00}\right. \\
& & \left.{}-{1\over 2}\sum_a m_a {{\bf x}_a}_i {\ddot{\bf x}_a}{}_j  {\bar h}_{ij}\right].
\end{eqnarray}
Adding together the result of Eq.~(\ref{eq:jeebus2}) plus its mirror image under $1\leftrightarrow 2$ and Eq.~(\ref{eq:jeebus3}) to Eq.~(\ref{eq:jeebus1}) we find the complete NRGR action at order $\sqrt{L} v^{5/2}$,
\begin{eqnarray}
\label{eq:NRGRv2}
L_{v^{5/2}} &=& -{{\bar h}_{00} \over 2 m_{Pl}}  \left[{1\over 2}\sum_a m_a {\bf v}^2_a -{G_N m_1 m_2\over |{\bf x}_1 -{\bf x}_2|}\right]\\
\nonumber
& &  -{1\over 2 m_{Pl}}\epsilon_{ijk} {\bf L}_k \partial_j {\bar h}_{i0} + {1\over 2 m_{Pl}}\sum_a m_a {{\bf x}_a}_i {{\bf x}_a}_j R_{0i0j}.
\end{eqnarray}
There are several things to notice about this equation.  First, we find that ${\bar h}_{00}$ couples at this order in $v$ to the Newtonian gravitational energy of the bound system.  At this order in the expansion this is a constant of the motion, as is the orbital angular momentum,  which couples to ${\bar h}_{i0}$.  Consequently, neither of these quantities can radiate.  This is in accord with the expectation that ${\bar h}_{00}, {\bar h}_{i0}$ do not correspond to physical graviton polarizations, and should not be emitted by a graviton source.  It should be possible to give a general diagrammatic proof that the diagrams in the matching involving one external ${\bar h}_{00}$ are related in a simple way to the diagrams with no external radiation gravitons, i.e, the diagrams that contribute to the total energy, as required by general covariance.

Also, note that this Lagrangian is manisfestly gauge invariant under infinitesimal gauge transformations.  This is obvious for the last term in Eq.~(\ref{eq:NRGRv2}), which involves the (linearized) Riemann tensor.  For the first term use the fact that the Newtonian energy is (at this order) time independent, so that under the transformation ${\bar h}_{00}\rightarrow {\bar h}_{00}+ 2\partial_0\xi_0$, this  term in the Lagrangian transforms into a total time derivative.  Likewise, under ${\bar h}_{i0}\rightarrow {\bar h}_{i0} +  \partial_i \xi_0 + \partial_0 \xi_i$, the coupling of ${\bar h}_{i0}$ to the angular momentum transforms into a total time derivative.  We expect that the requirement of gauge invariance should provide strong constraints on the form of the NRGR Lagrangian.  In particular, the derivation of the ${\cal O}(v^2)$ radiation couplings presented here could have been significantly shortened had we imposed gauge invariance.  In practice, one can reduce the amount of work by keeping only the terms from the multipole expansion and fixing the remaining terms by demanding gauge invariance of the radiation couplings.

It is interesting to note that in order for the couplings of the radiation gravitons to preserve gauge invariance, it is crucial that we include the diagrams of Fig.~\ref{fig:jeebus}b and Fig.~\ref{fig:jeebus}c.  These diagrams, which arise due to the non-Abelian nature of the gravitational interactions, represent the fact that in gravity it is not just the ensemble of NR particles that acts as a source of gravitons, but also the energy stored in the gravitational field that it sets up.  To obtain consistent results such non-linear effects must be included.   Without the velocity power counting rules of the EFT, it would have been difficult to ascertain whether such non-linear diagrams contribute already at this order in the velocity expansion.  In the classical general relativity approach to calculating the power emitted in gravitational waves by a NR source, this non-linearity appears in the expression for the far zone gravitational field as an expansion in moments not over the stress tensor $T^{\mu\nu}$ of the matter alone, but also over a suitable (non-unique) stress-energy ``pseudotensor" $\tau^{\mu\nu}$ for the gravitational field itself.   This pseudotensor is defined in such a way that $\partial_\mu T^{\mu\nu} +\partial_\mu \tau^{\mu\nu}=0$, and physically plays the same role as the diagrams of Fig.~\ref{fig:jeebus}b and Fig.~\ref{fig:jeebus}c do in our formalism.  As an aside, we also note that the necessity to include diagrams involving gauge particle self-interactions in order to derive the couplings of radiation to sources has an analog in the worldline approach to NR bound states in QCD~\cite{QCDworldline}.  For instance, to show that the color dipole moment of  a $Q\bar {Q}$ bound state couples to the full color electric field, it is not enough to keep diagrams with gluons emitted from the quark worldlines.  One also needs to keep diagrams involving the 3-gluon vertex, which can be shown to contribute at the same order in the velocity~\cite{peskin}.

\section{Divergences, RG Flows and Finite Size Effects}
\label{sec:RG}

\begin{figure*}[t!]
    \centering
    \includegraphics[width=0.9\textwidth]{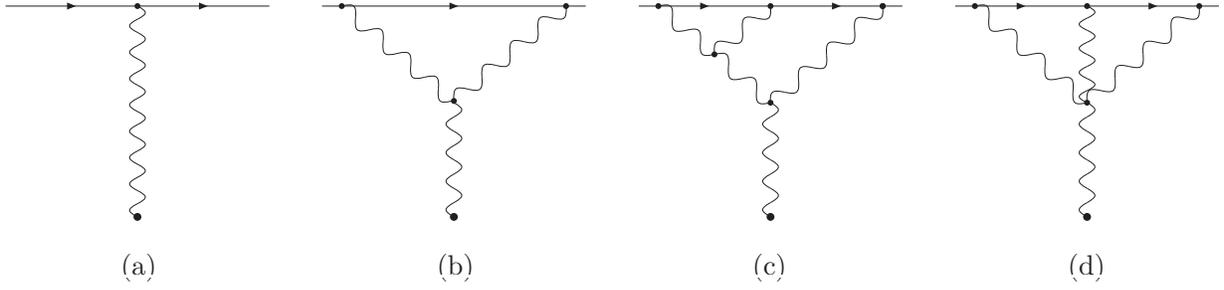}
\caption{Some contributions to the background graviton effective action in the relativistic point particle theory.  Diagram (b) has a power divergent term.  Diagrams (c) and (d) are both power and logarithmically divergent.  Two other diagrams that arise as permutations of (c) are not shown.}
\label{fig:tadpoles}
\end{figure*}

The coefficients of the relativistic point particle theory of Eq.~(\ref{eq:pp}) exhibit non-trivial RG flows, even at the classical level (RG flows in classical field theories coupled to extended spatial sources have been analyzed in~\cite{classicalRG}).  The non-trivial RG structure of the theory is associated with logarithmic UV divergences which arise due to the singular nature of the point particle limit.  To renormalize the theory we consider a single particle at rest at ${\bf x}=0$ and compute the diagrams that contribute to the terms in the effective action for a background gravitational field that are linear in the background $h_{\mu\nu}=g_{\mu\nu} - \eta_{\mu\nu}$.  If we ignore for now the non-minimal curvature couplings of Eq.~(\ref{eq:pp}), this is given by diagrams such as those of Fig.~\ref{fig:tadpoles} with a single external graviton and any number of insertions of the worldline proper time coupling, each proportional to the particle mass (we restrict ourselves to tree diagrams, which are the ones that remain in the limit $m_{Pl} r\gg 1$, with $r=|{\bf x}|$).  

While the diagram with a single mass insertion is finite, diagrams with more mass insertions have UV divergences.  For instance the diagram with two powers of $m$ in Fig.~\ref{fig:tadpoles}(b) contains a piece which has a power divergence of the form $\sim \int d^{d-1} {\bf q}/{\bf q}^2$, as can be seen by counting powers of momentum from the internal propagators and the graviton self-interaction vertex.  An explicit calculation shows that this linearly divergent piece, which vanishes in dimensional regularization, just renormalizes the bare mass parameter $m$ appearing in the point particle action.  Besides this power divergent piece, the diagram of Fig.~\ref{fig:tadpoles}(b) also gives rise to a finite, non-analytic term in the background field effective action~\cite{background} of the form
\begin{equation}
\Gamma[g_{\mu\nu}]={1\over m_{Pl}}\int  {d^4 k\over (2\pi)^4}h_{\mu\nu}(-k) T_{(2)}^{\mu\nu}(k)+\cdots,
\end{equation}
where, in background field gauge with a gauge fixing term for the ``quantum" graviton field $H_{\mu\nu}$ of the form $S_{GF}=-{1\over 2}\int d^4 x \sqrt{g} \Gamma_\mu \Gamma^\mu$, $\Gamma^\mu = D^\alpha H^\mu_\alpha -{1\over 2} D^\mu H^\alpha_\alpha$ one finds
\begin{eqnarray}
\nonumber
T_{(2)\mu\nu}(k) &=& (2\pi)\delta(k^0) {m^2\over 2 m^2_{Pl}}\left[-{7\over 32}(\eta_{\mu\nu} k^2 -k_\mu k_\nu)\right. \\
\nonumber
& &\left. {} + {1\over 32} k^2 v_\mu v_\nu\right]\int {d^3{\bf q}\over (2\pi)^3} {1\over {\bf q}^2 ({\bf q} +{\bf k})^2} \\
\nonumber
&=& (2\pi)\delta(k^0) {m^2\over 16 m^2_{Pl} |{\bf k}|}\left[-{7\over 32}(\eta_{\mu\nu} k^2 -k_\mu k_\nu)\right.\\
& & {}\left.  +  {1\over 32} k^2 v_\mu v_\nu\right]
\end{eqnarray}
where $v^\mu=(1,{\bf 0})$ is the particle four-velocity.  As expected by background field gauge invariance, we have $k_\mu T^{\mu\nu}_{(2)}(k)=0$.  In fact, since $\Gamma[g_{\mu\nu}]$ must be covariant with respect to the background gravitational field, the above result is sufficient to completely fix all terms in the effective action that are proportional to two powers of $m$ and any number of powers of the background field $h_{\mu\nu}$.

Although the power divergences that arise in the calculation of $T^{\mu\nu}_{(2)}(k)$ have no physical consequences, logarithmic divergences can give rise to measurable effects since by dimensional analysis, in a massless field theory the logarithm of the cutoff is always accompanied by a logarithm of the external momentum.   Consider the diagrams with three insertions of the particle mass.  They give rise to a term in the effective action of the form
\begin{equation}
\Gamma[g_{\mu\nu}]={1\over m_{Pl}}\int {d^4 k\over (2\pi)^4}h_{\mu\nu}(-k) T_{(3)}^{\mu\nu}(k)+\cdots,
\end{equation}
where we find, from Fig.~\ref{fig:tadpoles}(c) and Fig.~\ref{fig:tadpoles}(d) 
\begin{widetext}
\begin{equation}
\label{eq:T3}
T_{(3)}^{\mu\nu}(k)=
 {1\over 3!} \left(- {im\over 2 m_{Pl}}\right)^3 {2\over m_{Pl}}(2\pi)\delta(k^0)I_0({\bf k})\left[{1\over 16}(1-2\epsilon)\left(k^2\eta^{\mu\nu} - k^\mu k^\nu\right) -{1\over 8}\left(1-{21\over 2}\epsilon\right)k^2 v^{\mu} v^{\nu}\right].
\end{equation}
\end{widetext}
As a non-trivial check of this result, we note that gauge invariance with respect to the background field, $k_\mu T_{(3)}^{\mu\nu}(k)=0$, does not hold diagram by diagram, but is only manifest after the terms in Fig.~\ref{fig:tadpoles}(c) and Fig.~\ref{fig:tadpoles}(d) are combined.  To obtain this equation, we worked out the necessary background field gauge Feynman rules \cite{veltman} in $d$ dimensions, and performed an expansion in $\epsilon=(4-d)/2$, keeping only the terms that do not vanish as $\epsilon\rightarrow 0$.  $T^{\mu\nu}_{(3)}(k)$ contains a UV logarithmic divergence, as can be seen by power counting the integral 
\begin{equation}
I_0({\bf k}) = \int {d^{d-1}{\bf p}\over (2\pi)^{d-1}} {d^{d-1}{\bf q}\over (2\pi)^{d-1}} {1\over {\bf q}^2 {\bf p}^2 ({\bf q} + {\bf p} + {\bf k})^2},
\end{equation}
 which evaluates to
\begin{eqnarray}
\nonumber
I_0({\bf k}) &=& \sqrt{\pi}{\Gamma(4-d)\over (4\pi)^{d-1}} {\Gamma(d/2-3/2)^2\Gamma(d-3)\over \Gamma(d/2-1)\Gamma(3d/2 - 9/2)}   \left({{\bf k}^2\over 2}\right)^{d-4}\\
\nonumber
&=& {1\over 32\pi^2}\left[{1\over 2\epsilon} +\ln (4\pi) -\gamma +3 -\ln{{\bf k}^2\over\mu^2} \right]+{\cal O}(\epsilon).\\
\end{eqnarray}
In the second line we have expanded about $d=4-2\epsilon$ and introduced an arbitrary subtraction scale $\mu$.  Inserting this expression back into Eq.~(\ref{eq:T3}), we see that the set of diagram with three mass insertions has a $1/\epsilon$ pole which cannot be absorbed by a shift in the mass parameter $m$ of the minimal point particle action, whose Feynman rule is, for a particle moving on a straight line trajectory with constant four-velocity $v^\mu$,
\begin{equation}
-{im\over 2 m_{Pl}} (2\pi)\delta(k\cdot v) v^{\mu} v^{\nu}.
\end{equation}
However, the non-minimal operators of Eq.~(\ref{eq:pp}) contribute terms to $\Gamma[g_{\mu\nu}]$ of the form $(1/m_{Pl})\int d^4 k/(2\pi)^4 h_{\mu\nu}(-k) T_{ct}^{\mu\nu}(k)$, with
\begin{equation}
T_{ct}^{\mu\nu}(k)= (2\pi) \delta(k^0)\left[c_R (\eta^{\mu\nu} k^2 - k^\mu k^\nu) +{1\over 2} c_V k^2 v^\mu v^\nu\right],
\end{equation}
where we have dropped terms proportional to $k^0$ which vanish due to the delta function.  Because the divergent terms in Eq.~(\ref{eq:T3}) have the same dependence on $k$ and $v$ as the contribution from the curvature couplings, we see that the $1/\epsilon$ poles can be subtracted away by a proper choice of coefficients $c_{R,V}$.  In the $\overline{MS}$ scheme, the renormalized ${\cal O}(m^3 h)$ term in the effective action now becomes
\begin{widetext}
\begin{equation}
\label{eq:renormT3}
T_{(3)}^{\mu\nu}(k) = (2\pi) \delta(k^0)\left[(\eta^{\mu\nu} k^2 -k^\mu k^\nu)\left\{c_R(\mu) + {G_N^2 m^3\over 12}\left(2-\ln{{\bf k}^2\over \mu^2}\right)\right\} - k^2 v^\mu v^\nu\left\{-{1\over 2} c_V(\mu) +{G_N^2 m^3\over 6}\left({75\over 32}-\ln{{\bf k}^2\over\mu^2}\right)\right\}\right],
\end{equation} 
\end{widetext}
where $c_{R,V}(\mu)$ are the renormalized couplings.  

Since the renormalization scale $\mu$ introduced above is arbitrary, we must have $\mu d\Gamma[g_{\mu\nu}]/d\mu=0$ (since the classical field $h_{\mu\nu}$ does not pick up an anomalous dimension).  Thus the explicit dependence of the logarithm in Eq.~(\ref{eq:renormT3}) on the subtraction scale $\mu$ must be cancelled by allowing the coefficients $c_{R,V}$ to vary with scale.  Our theory therefore exhibits non-trivial {\it classical}  RG scaling,
\begin{eqnarray}
\label{eq:RGcR}
\mu {d c_R\over d\mu} &=& -{1\over 6} G_N^2 m^3, \\
\label{eq:RGcV}
 \mu {d c_V\over d\mu} &=& {2\over 3} G_N^2 m^3.
 \end{eqnarray}

As a consistency check, we may use the linear term in $\Gamma[g_{\mu\nu}]$ to calculate the spacetime metric generated by a point mass at rest, up to order $(G_N m/r)^3$.  Fixing the gauge $\eta^{\mu\nu} \Gamma^\lambda_{\mu\nu}=0$ for the background field $h_{\mu\nu}$, this is simply given by $g_{\mu\nu}=\eta_{\mu\nu} + h_{\mu\nu}$, where, in momentum space
\begin{equation}
h_{\mu\nu}(k) = {i\over k^2} P_{\mu\nu;\alpha\beta}  \left.{i\over m_{Pl}} {\delta \Gamma[g]\over \delta h_{\alpha\beta}(-k)}\right|_{h=0}. 
\end{equation} 
Using our results for $T^{\mu\nu}_{(i)}(k)$, we find, dropping delta function contributions which first arise at order $m^3$,
\begin{widetext}
\begin{eqnarray}
g_{00}  &=& 1 -  {2G_N m\over r} + 2  \left({G_N m\over r}\right)^2  + 2\left( {G_N m\over r}\right)^3+\cdots,\\
\nonumber
g_{ij}   &=& -\delta_{ij}\left[1 + {2 G_N m\over r} + 5 \left( {G_N m\over r}\right)^2-{2\over 3} \left( {G_N m\over r}\right)^3+ 8 \left( {G_N m\over r}\right)^3\left\{{c_R(\mu)\over G_N^2 m^3} + {1\over 6}\ln\mu r +{\gamma\over 6}\right\}\right]\\
& & {}  + {x_i x_j\over r^2}\left[7\left({G_N m\over r}\right)^2 -{4\over 3} \left( {G_N m\over r}\right)^3+ 24 \left( {G_N m\over r}\right)^3\left\{{c_R(\mu)\over G_N^2 m^3} + {1\over 6}\ln\mu r +{\gamma\over 6}\right\}\right]+\cdots ,
\end{eqnarray}
\end{widetext} 
and $g_{0i}=0.$  It is straightforward to show that this metric satistfies the vacuum Einstein equations $R_{\mu\nu}=0$ away from the point ${\bf x}=0$ where the point particle source is located, independently of the value of the renormalized parameter $c_R(\mu)$.  This is consistent with Birkhoff's theorem, which states that the space of static, rotation invariant solutions of the vacuum Einstein equations is one-dimensional, parametrized by the mass $m$.  Alternatively, one may check the validity of our result for $g_{\mu\nu}$ by constructing a coordinate transformation to the more familiar harmonic coordinates ${\bar x}^\mu,$ with ${\bar g}^{\mu\nu}({\bar x})\Gamma^{\lambda}_{\mu\nu}({\bar x})=0$, which in terms of our coordinates $x^\mu$ are given by, up to ${\cal O}(G_N^4 m^4/r^4)$ terms,
\begin{eqnarray}
{\bar x}^0 &=& x^0,\\
{\bar x}^i &=& x^i\left[1+2 \left({G_N m\over r}\right)^2 -{7\over 3}\left({G_N m\over r}\right)^3\right.\\
\nonumber
& & \left. {} + 4 \left({G_N m\over r}\right)^3\left\{{c_R(\mu)\over G^2_N m^3}+{1\over 6}\ln\mu r +{\gamma\over 6}\right\}\right].
\end{eqnarray}
Thus, as required by general arguments, the parameters $c_{R,V}(\mu)$ are not observable in the long range field of isolated static sources.  

As we argue in the next section, it is likely that the value of the matching coefficients  $c_{R,V}(\mu_0)$ are themselves of order $G_N^2 m^3$.  In this case, it is easy to see using the NRGR power counting rules that diagrams with one insertion of the non-minimal operators in Eq.~(\ref{eq:pp}), for instance the diagram of  Fig.~\ref{fig:v6logv1}(a) lead to corrections to the  two-body interaction Lagrangian which are suppressed by a power of $v^6$ relative to the leading Newton interaction.  However, it can be shown that the worldline operators $\int d\tau R, \int d\tau R_{\mu\nu} {\dot x}^\mu {\dot x}^\nu$ can be removed by a local redefinition of the metric tensor (that is, the coefficients $c_{R,V}$ can be shifted arbitrarily by these transformations).   Given that any possible signature of internal structure that may arise at $v^6$ must be encoded in these two operators withing a point particle description, we conclude for spinless objects {\it there are no finite size effects up to order $v^6$}.

 Even though all values of $\mu$ give rise to the same physics, in practice certain choices may be more convenient when doing actual computations.  In general, this leads to renormalization group equations which can be used to simplify the calculation of logarithmically enhanced terms in the perturbative expansion.   We will give an example in Section~\ref{sec:v6log} of how to exploit the freedom in choosing $\mu$ for the unphysical couplings $c_{V,R}$ to simplify the calculation of certain contributions to the $v^6$ two-body potentials that are enhanced by a logarithm of $\alpha=r_s/r$, where $r_s$ is a length scale that characterizes the size of the extended object.  The calculation of logarithms for physical tidal couplings, which involve the Riemann tensor, is completely analogous to this case. 
  
Any choice for the couplings $c_{R,V}(\mu)$ can be trivially expressed in terms of the parameters at some fixed scale $\mu_0$ by solving the above RG equations  
 \begin{eqnarray}
 \label{eq:RGsol}
 c_R(\mu)&=& c_R(\mu_0)-{1\over 6} G_N^2 m^3\ln{\mu\over\mu_0},\nonumber \\
 c_V(\mu)&=& c_V(\mu_0) + {1\over 3} G_N^2 m^3\ln{\mu\over \mu_0}.
 \end{eqnarray}
 (as we will see later, at higher order in the velocity expansion, the coupling constant flows show more interesting behavior than is depicted in these equations).   If the couplings $c_{R,V}(\mu_0)$ had any physical content, in order to fix them say at a scale of order the characteristic size $r_s$, requires model-dependent information about the details of the short wavelength physics that resolves the point particle divergences.  This will be true for the physical tidal couplings as well.   This physics is independent of the NR two-body dynamics, so that no PN formalism based on point particles can, by itself, give rise to a full prediction for the gravitational wave observables.  In the next section we will explain how, given this short distance model, it is possible to perform a matching calculation that fixes the values of  the worldline operator coefficients in terms of the parameters (size, equation of state, $\ldots$) that characterize the extended object.
 
An interesting question arises as to the size of the matching coefficient $c_{R,V}(\mu_0)$, or their more physical tidal counterparts.  If it turned out that these coefficients were anomalously large\footnote{An explicit example of this arises in some models of branes embedded in higher-dimensional spacetimes~\cite{DGP}.}, then this would put the systematics of the PN approximation in jeopardy.  For instance if the coefficients in Eq.~(\ref{eq:pp}) were much larger than $G_N^2 m^3$ (no explicit powers of $v$, a dynamical variable, can appear in operator coefficients), the Newtonian result would no longer be leading order, and the PN expansion would cease to be valid.  Although it seems rather unlikely that this constraint is violated, it is interesting to note that within the EFT formalism here, we would retain calculability given a model for the short distance physics, but we would have to change our power counting scheme in order to accomodate the lack of a Newtonian gravity limit.

\section{Matching on to the Point Particle Theory}
\label{sec:matching}

\subsection{Outline of the Matching Procedure}
If one works at high enough of an order for the worldline operators to become relevant, i.e. beyond $v^6$, then given that PN theory has nothing to say about the actual values of the worldline operator coefficients, one needs a model of the internal structure. Otherwise  the best one can do is to express the predictions for the gravitational wave observables in terms of the parameters of the EFT and use the gravitational wave data to obtain information about the magnitude of the coefficients.  Alternatively, if one does have information about the structure of the binary constituents, one can fix the values of the couplings in the effective action at a renormalization point $\mu\sim 1/r_s$.  In this case the data can, in principle,  be used to test one's assumptions regarding the structure of the binary star sources.  In this paper we will not perform the matching calculation, but instead just sketch how such a calculation would lead to an unambiguous prediction for the gravitational wave observables.   We will illustrate how this works within the context of a specific toy model which captures the essential physics.  
 
The matching procedure is standard within effective theories~\cite{EFTrev}, but to keep the discussion self-contained we will review here the basic concepts.  The idea is simply to compare observables such as Green's functions in the ``full" (valid at short distances) and effective theories and adjust the value of the EFT couplings in such a way that the effective theory reproduces the results of the full theory.  The effective theory is designed to reproduce the long distance physics, characterized by non-analytic behavior about zero external momentum, but it may get the short distance, which is analytic in the momentum (that is, local in coordinate space), wrong (hence the divergences).
 
 Consider for instance the matching of one-point functions in the full theory (consisting of an extended source interacting with gravitons) and the EFT.  In the full theory, the one point-function of the graviton are of the form
\begin{equation}
G^{(1)}(k^2) = {1\over k^2} F(k^2,{\kappa_i}),
\end{equation}
where $F(k^2,{\kappa}_i)$ is in general a complicated function of the external momentum $k$ as well as some dimensionful parameters $\kappa_i$ which describe internal structure.  For simplicity we assume in the discussion that the only relevant scale is the size of the object $r_s$.      The EFT is valid in the kinematic regime $k^2 r^2_s \ll 1$, so in order to match, one expands the function $F(k^2 r_s^2)$ about this point.  In the simplest situations, this expansion takes the form 
 \begin{equation}
 \label{eq:gfull}
G^{(1)}(k^2) = {1\over k^2}\left[P(k^2 r_s^2) + Q(k^2 r_s^2)\ln(k^2 r_s^2)\right],
 \end{equation}
 where $P(z),Q(z)$ are polynomials.  In general, the non-analytic dependence on the momentum may be more complicated, involving polynomials times more general functions of $\ln k^2 r_s^2$, however, we focus on this case for simplicity.  In practice, since we are interested in a low energy expansion, we need to keep only the first few terms in the polynomials $P(z),Q(z)$.   Provided that the point particle coupled to gravity properly reproduces the IR (long distance) physics, the EFT result must be of the same general form as the above 
 \begin{equation}
 \label{eq:gEFT}
G^{(1)}_{EFT}(k^2) = {1\over k^2}\left[{\bar P}(k^2,c_i(\mu)) + {\bar Q}(k^2,c_i(\mu))\ln {k^2\over\mu^2}\right],
 \end{equation}
where the parameters $c_i(\mu)$ are the coupling constants of the EFT defined by some suitable renormalization prescription, which in the point particle theory are the particle mass and the curvature couplings.  The functions ${\bar P},{\bar Q}$ are polynomials of $k^2$ whose coefficients are products of the $c_{i}(\mu)$.  Note that the log divergence arises because we have taken the singular limit $r_s\rightarrow 0$.  Since this logarithm arises in the full theory from the light degrees of freedom which the EFT must also contain, it is crucial that the non-analytic momentum dependence in Eq.~(\ref{eq:gEFT}) coincides with that of Eq.~(\ref{eq:gfull}).  In order for the EFT to reproduce the results of the full theory (at least to some fixed order in the small parameter $k^2 r_s^2$), we then tune the coefficients $c_{R,V}(\mu)$ so that $G^{(1)}(k^2)=G^{(1)}_{EFT}(k^2)$ in the limit $k^2 r_s^2\ll 1$.  

For instance, in the previous section, we had ${\bar P}\sim m/m_{Pl}+c_{R,V}(\mu)k^2/m_{Pl},$ and ${\bar Q}\sim G_N^2 m^3 k^2/m_{Pl}$, hence in the full theory  that resolves the point particle singularities, we expect $Q(z)\sim G^2_N m^3 z/r_s^2 m_{Pl}$ for $z\rightarrow 0$ (we are ignoring finite, non-analytic power corrections of the form $\sqrt{k^2}$ from Fig~\ref{fig:tadpoles}b, which are not relevant to this discussion).  Comparing Eq.~(\ref{eq:gfull}) and Eq.~(\ref{eq:gEFT}) in this case we obtain a relation
\begin{equation}
c_{R,V}(\mu) \sim  c_0+ G_N^2 m^3 \ln(\mu r_s),
\end{equation}      
where $c_0$ is a full theory parameter obtained from the expansion of $P(z)$ to linear order.  To avoid large logs which could hinder our perturbative expansion, we choose $\mu\sim1/r_s$.  This is the boundary value of the solution to the RG equation Eq.~(\ref{eq:RGsol}).  Actually, this equation ignores the presence of scheme dependent constants (which are not relevant to physical predictions) which are of the same magnitude as the coefficient of the logarithm.  Thus even if the short distance parameter $c_0$ were much smaller than $G^2_N m^3$, we would find $c_{R,V}(\mu_0)\sim G_N^2 m^3$.

\subsection{A Toy Model of the  Short Distance Physics}

In this section we illustrate the general remarks above in the context of two simple toy models.  The matching of the worldline operator coefficients  which carry information about the static properties is conceptually similar to the matching in a model consisting of a self-interacting scalar field $\phi$ coupled to a point-particle source.  To match point particle couplings which carry dynamical information about the binary star constituents, for instance a tidal operator of the form $\int d\tau R_{\mu\nu\alpha\beta}  R^{\mu\nu\alpha\beta}$ that may appear in the point particle EFT, requires matching amplitudes with more than one external graviton.  We will consider an analog of this in electrodynamics, where we will describe the matching onto the point particle operators $\int d\tau F_{\mu\nu} F^{\mu\nu}$,  $\int d\tau {\dot x}^\mu {\dot x}^\nu F_{\mu\alpha} F^{\alpha\nu}$ which encode the dynamical repsonse of a charge distribution to applied low frequency electromagnetic fields.  Explicit calculations of the counterterms need to fix the operators coefficients for the point particle EFT a gravitating compact object are currently in progress~\cite{us}.

First  consider a self-interacting massless scalar field $\phi(x)$ in four dimensions coupled to a localized source $\rho(x)$, 
\begin{equation}
\label{eq:UVmodel}
{\cal L} = {1\over 2} (\partial \phi)^2 - {\lambda\over 4!}\phi^4 -\rho(x) \phi(x),
\end{equation}
and imagine that through some unspecified dynamics, there is a frame in which the source $\rho(x)$ acquires a profile of the form
\begin{equation}
\rho(x^0,{\bf x}) = {{\bar g}\over \pi^{3/2} r^3_s}\exp\left[-{\bf x}^2/r^2_s\right],
\end{equation}
where ${\bar g}$, and $r_s$ are constants that parametrize the short distance physics.  Although this model is unrealistic in the sense that the profile $\rho(x)$ does not arise dynamically but is rather put in by hand, it is adequate for our purposes here.  

In the limit where only long wavelength modes of the field $\phi(x)$ are accessible, the detailed structure of the source $\rho(x)$ becomes irrelevant.  As far as long distance modes are concerned, one may replace the localized source with a point particle that has suitable local couplings to the field $\phi(x)$.  
\begin{equation}
S_{eff} = S[\phi] - m\int d\tau -g \int d\tau \phi +\cdots,
\end{equation}
where $S[\phi]$ is the bulk scalar field action, and we have suppressed higher dimension worldline couplings involving more powers of $\phi$ or its derivatives.  The coupling $g$ appearing in this equation is a function of the paramaters ${\bar g}$, $r_s$ of the short distance source model, which we determine by matching amplitudes in the low energy EFT with those obtained from Eq.~(\ref{eq:UVmodel}).  At lowest order in $\lambda$, this is trivial.  Consider for instance the amplitude for $\mbox{vacuum}\rightarrow\phi$.  In the full theory
\begin{equation}
i(2\pi)\delta(k^0){\cal A}_{full} = -i\int d^4 x \rho(x) \langle k|\phi(x)|0 \rangle\\
\end{equation}
so that 
\begin{equation}
i{\cal A}_{full} = - i{\bar g} \exp\left[-{\bf k}^2 r^2_s/4\right],
\end{equation}
which essentially is the form factor associated with the source distribution $\rho(x)$.  In the EFT,
\begin{eqnarray}
\nonumber
i(2\pi)\delta(k^0){\cal A}_{EFT} &=& -ig\int d\tau \langle k|\phi(x)|0 \rangle\\
&=& - i(2\pi)\delta(k^0) g. 
\end{eqnarray}
The two amplitudes agree in the limit ${\bf k}^2 r^2_s\rightarrow 0$ if we set $g={\bar g}.$  At higher order in the momentum expansion we would also need operators of the form $\int d\tau\partial^{2n}\phi$ in the low energy theory to reproduce the full theory amplitude.  Similarly, by matching the vacuum to vacuum amplitudes, we find $m=(1/4\sqrt{2\pi}){\bar g}^2/r_s$.  

Note that in the point particle EFT, there are logarithmic divergences in the $\mbox{vacuum}\rightarrow\phi$ amplitude at ${\cal O}(g^3\lambda)$, from a diagram analogous to the one in Fig.~\ref{fig:tadpoles}(d),
\begin{eqnarray}
i{\cal A}_{EFT} &=& -ig + {i\lambda g^3\over 3!} I_0({\bf k})\\
\nonumber
 &=& -ig(\mu) +{i\lambda g^3\over 192\pi^2}\left[3-\ln{{\bf k}^2\over\mu^2}\right],
\end{eqnarray}    
where $I_0({\bf k})$ was defined in the previous section, and as before we have used dimensional regularization plus the $\overline{MS}$ scheme to renormalize the divergent amplitude.  If the EFT is to be a correct description of the low energy physics, the non-analytic dependence on the momentum in this equation should reproduce a similar log in the full theory amplitude.  Indeed, calculating in the full theory and including insertions of the $\phi^4$ interaction we find, dropping terms subleading in ${\bf k}^2 r^2_s\ll 1$
\begin{eqnarray}
i{\cal A}_{full} &\simeq&-i{\bar g} +  {i\lambda {\bar g}^3\over 48}\int {d^3{\bf q}\over (2\pi)^3} {e^{-{\bf q}^2 r_s^2/4} \over {\bf q}^2 |{\bf q} + {\bf k}|}\\
\nonumber
&\rightarrow&-i{\bar g} - {i\lambda{\bar g}^3\over 192\pi^2}\left(\ln{{\bf k}^2 r^2_s\over 4} + \gamma -1\right).
\end{eqnarray}
which agrees with the claimed universal result Eq.~(\ref{eq:gfull}).
Therefore
\begin{equation}
\label{eq:UVmatch}
g(\mu)  = {\bar g}  + {\lambda{\bar g}^3 \over 192\pi^2}\left[\ln{\mu^2 r^2_s\over 4} +\gamma +2\right],
\end{equation}
which indicates that to avoid large logarithms that may render perturbation theory invalid, we should match at a scale $\mu\sim 1/r_s$, so that $g(1/r_s)\simeq {\bar g}$.  Then we can use the RG equation
\begin{equation}
 \mu {dg\over d\mu}   = {\lambda g^3\over 96\pi^2}
\end{equation}
 to sum logarithms of ${\bf k}^2 r_s^2$.  Even though we phrased the matching problem using quantum field theory language, we stress that our discussion here applies also in the classical regime.  This can be seen by matching not scattering amplitudes, but the 1-point functions $\langle\phi(x)\rangle$, in the presence of the point particle source, which have an obvious classical interpretation.
 
How does this model bear on the matching problem for the point particle coupled to gravity?  As in the gravitational case, our toy scalar field model exhibits UV divergences in the point particle limit.  In both cases, we expect those divergences to be resolved by the finite extent of the distribution which sources the field.  We saw this explicitly in the scalar field theory, where we learned that the matching procedure fixes the couplings of the point particle at a length scale of order the size of the source.  We expect that the same will occur in the gravitational problem.  

However, there are several aspects of the gravity problem that are not captured by this simple model.  In the scalar field theory, the UV divergences associated with the point particle singularities affect the leading order coupling $\int d\tau\phi$ itself, and thus the point $g=0$ is a (trivial) fixed point in the RG sense.  In the gravitational case, on the other hand, the structure of the RG flow is such that even if the curvature couplings were equal to zero at $\mu\sim 1/r_s$, non-zero values would get induced at longer wavelengths.  Second, in the scalar model, the order $\lambda g^3$ correction to the long range field encodes information about the internal structure of the extended object that couples to $\phi$.  For a source in isolation coupled to gravity, Birkhoff's theorem states that the curvature couplings cannot generate long range static forces that could carry information about the short distance dynamics.  Nevertheless, Birkhoff's theorem does not prohibit the momentum space off-shell graviton one-point  function from having sensitivity to the internal structure.  Thus in principle one could still match the graviton one-point functions (or the quantities $T_{(i)}^{\mu\nu}(k)$ appearing in the background field action),  although in practice it may be difficult to compare in a gauge invariant manner.  Because of this, it may be simpler to match by comparing gauge invariant quantities, such as a the amplitudes for fields to scatter off the geometry of the extended source versus the scattering amplitude in the point particle theory.

Fixing the coefficients of the operators that measure tidal effects in the point particle EFT will require matching amplitudes with more external gravitons.  A similar situation arises in electrodynamics, where dynamical effects such as induced dipole moments are encoded in the photon two-point function (equivalently the response to an external field).  Consider for instance the effective point particle action that describes the interactions of the electromagnetic field with a perfectly conducting grounded sphere of radius $r_s$.  Since by definition the times scales associated with the dynamics of a perfect conductor  are short compared to those of the electromagnetic fields it interacts with, the electrodynamic analog of Eq.~(\ref{eq:pp}) is then
\begin{eqnarray}
S_{pp}  &=& -m\int d\tau + eQ \int dx^\mu A_\mu\\
\nonumber
& & {}  + {\alpha\over 2}\int d\tau F_{\mu\nu}F^{\mu\nu} + {\beta\over 2}\int d\tau {\dot x}^\mu {\dot x}^\nu F_{\mu\sigma} F^\sigma_\nu.
\end{eqnarray} 
Imagine putting this system in a constant background electromagnetic field ${\cal F}_{\mu\nu}$. 
Expanding the field about $\cal{F}$ in the point particle EFT, the fluctuating photon field $A_\mu$ acquires a vacuum expectation value which is given, in the Feynman gauge, by
\begin{equation}
\langle A_\mu(x)\rangle = {e Q\over 4\pi|{\bf x}|}\eta_{\mu 0} - {\alpha\over 2\pi} {{\bf x}^i\over |{\bf x}|^3} {\cal F}_{\mu i} + {\beta\over 4\pi}\eta_{\mu0} {{\bf x}^i\over |{\bf x}|^3} {\cal F}_{0i}.
\end{equation}
Imposing the appropriate boundary conditions for a perfect conductor, the ``microscopic" model predicts
\begin{equation}
\langle A_\mu(x)\rangle = -{r^3_s\over 2}{{\bf x}^i\over |{\bf x}|^3}{\cal F}_{\mu i} + {3r^3_s\over 2}\eta_{\mu0} {{\bf x}^i\over |{\bf x}|^3} {\cal F}_{0i},
\end{equation}
so matching fixes the gauge invariant EFT coefficients to be $Q=0$, $\beta=6\alpha = 6\pi r^3_s.$  In this particular model, all higher dimension operators built out of more derivatives or powers of $F_{\mu\nu}$ have vanishing coefficients.

\section{Logarithmically Enhanced Potentials}
\label{sec:v6log}

Although the operators $\int d\tau R$ and $\int d\tau {\dot x}^\mu {\dot x}^\nu R_{\mu\nu}$ can be removed by a redefinition of the metric and thus never contribute to physical observables, it is instructive to use these operators to illustrate the general procedure for calculating the contributions to the two-body potential which are enhanced by a logarithm of $\alpha=r_s/r$.  Such terms are independent of the details of the short distace dynamics.  Although these terms have been obtained in a different gauge by conventional PN methods in Refs.~\cite{PNv6pot}, here we give a derivation that exploits the RG scaling of the coefficients in the EFT to simplify the calculations.  

The idea is that because the amplitudes in the EFT are independent of the subtraction scale $\mu$ we can choose any convenient value for it.  In particular, if one chooses $\mu$ to be of order the orbital scale $1/r$, then the terms in the two-body potential that involve logarithms of $\mu |{\bf x}_{12}|$ (which arise from taking the Fourier transform of diagrams involving subgraphs such as those in Fig.~\ref{fig:tadpoles}c,d) are minimized.  In this case the log enhanced contribution to any amplitude can be obtained entirely from diagrams with insertions of the local operators $\int d\tau R$,  $\int d\tau \dot{x}^\mu \dot {x}^\nu R_{\mu\nu}$, with coefficients $c_{R,V}$ evaluated at a scale $\mu\sim 1/|{\bf x}_{12}|$.  This procedure amounts essentially to removing the short distance modes of the gravitational field with wavelengths between the internal distance $r_s$ and the orbital scale $r$.

\begin{figure}
    \centering
    \includegraphics[width=4cm]{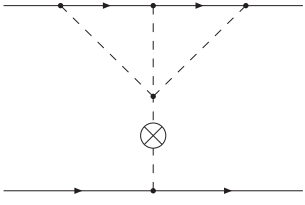}
\caption{Sample contribution to two-body potentials at $v^6\ln\alpha$.  All particle vertices are the leading order mass $v^0$ mass insertion.  The $\otimes$ denotes an insertion of the potential graviton kinetic term.}
\label{fig:v6small}
\end{figure}

For example, consider the diagram shown in Fig.~\ref{fig:v6small}, which contributes to the $v^6\ln\alpha$ terms in the potential.  Schematically, it contains terms of the form
\begin{eqnarray}
\mbox{Fig.~\ref{fig:v6small}}\sim \int dx^0{G_N m_2v^2\over |{\bf x}_1 - {\bf x}_2|^3}\left[c(\mu) + G^2_N m_1^3\ln(\mu|{\bf x}|_{12})\right],
\end{eqnarray}
where $c(\mu)$ denotes the contributions that arise due to insertions of the non-minimal curvature couplings and the logarithmic piece arises from a calculation similar to that of the ${\cal O}(m^3)$ terms of the background graviton effective action in Section~\ref{sec:RG}.  Although it is hard in general to calculate those logarithmic pieces, their argument is invariably of the form $\mu |{\bf x}_{12}|$, so by choosing $\mu\sim 1/|{\bf x}_{12}|$ we may eliminate these pieces in favor of the effective couplings $c(1/|{\bf x}_{12}|)$.   These in turn may be obtained in terms of the bare parameters $c(1/r_s)$ by using the RG equations for the couplings in $c_{R,V}$ in Eq.~(\ref{eq:RGcR}), Eq.~(\ref{eq:RGcV}),
\begin{equation}
\nonumber
c(1/r)\sim c(1/r_s) + G^2_N m^3 \ln\alpha. 
\end{equation}
Since the validity of the velocity expansion requires $c(1/r_s)\leq G_N^2 m^3$, the logarithmic term gives the dominant contribution to the two-body potential at this order in $v$.  
\begin{figure*}[t]
    \centering
    \includegraphics[width=0.9\textwidth]{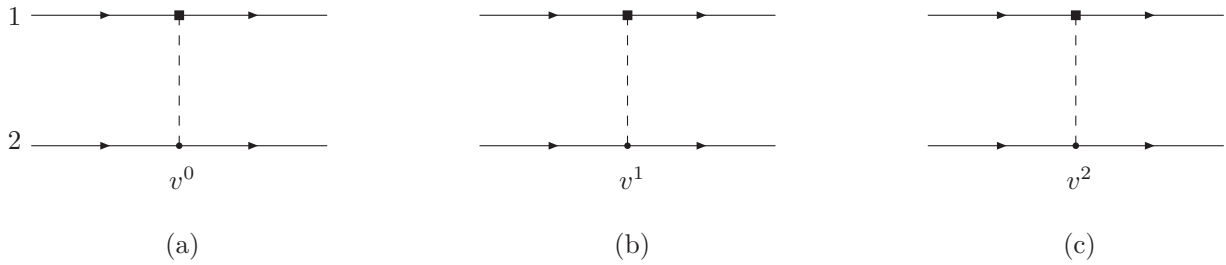}
\caption{One potential graviton exchange diagrams contributing to the two-body Lagrangian at $v^6\ln\alpha$.  The heavy dot indicates an insertion of the $\int d\tau R$ or $\int d\tau R_{\mu\nu} {\dot x}^\mu {\dot x}^\nu$ vertex with $c_{R,V}(\mu)$ evaluated at $\mu\sim 1/r$.  Similar diagrams with $1\leftrightarrow 2$ not shown.}
\label{fig:v6logv1}
\end{figure*}

\begin{figure*}[t]
    \centering
    \includegraphics[width=0.9\textwidth]{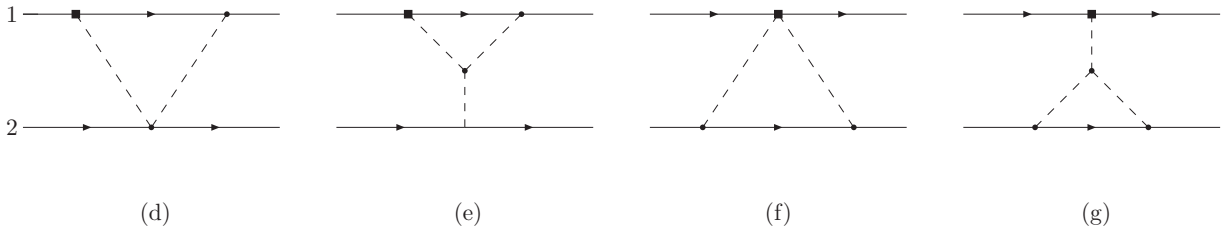}
\caption{Two and three potential graviton exchange diagrams contributing to the two-body Lagrangian at $v^6\ln\alpha$.  Similar diagrams with $1\leftrightarrow 2$ not shown.}
\label{fig:v6logv2}
\end{figure*}

Following this line of reasoning, we can obtain the model independent $v^6\ln\alpha$ terms by calculating the diagrams of Fig.~\ref{fig:v6logv1}, Fig.~\ref{fig:v6logv2} and evaluating the coefficients $c_{R,V}(\mu)$ at the scale $\mu\sim 1/|{\bf x}_{12}|$.  We find
\begin{widetext}
\begin{eqnarray}
\label{eq:v6logpot} 
\nonumber
L_{v^6\ln\alpha} &=& - 4 G_N \left(m_2 c_R^{(1)}(|{\bf x}_{12}|) + m_1  c_R^{(2)}(|{\bf x}_{12}|\right)\left[ {{\bf v}_1 \cdot {\bf v}_2\over |{\bf x}_{12}|^3} - {3 ({\bf v}_1\cdot {\bf x}_{12}) ({\bf v}_2\cdot {\bf x}_{12})\over |{\bf x}_{12}|^5}\right]\\
& & {} + \left\{4 G_N m_2 c^{(1)}_R(|{\bf x}_{12}|)\left[{{\bf v}^2_2\over |{\bf x}_{12}|^3}- {3 ({\bf v}_2\cdot {\bf x}_{12})^2 \over |{\bf x}_{12}|^5}\right] + (1\leftrightarrow 2)\right\} - {4G^2_N m_1 m_2\over |{\bf x}_{12}|^4} \left(c^{(1)}_R(|{\bf x}_{12}|) + c^{(2)}_R(|{\bf x}_{12}|)\right). 
\end{eqnarray}
\end{widetext}
In this equation, the first and second terms follow from the diagrams with a single graviton exchange in Fig.~\ref{fig:v6logv1}, while the last term comes from the diagrams with two and three graviton propagators, Fig.~\ref{fig:v6logv2}.  For the diagrams involving insertions of the operator $\int d\tau R_{\mu\nu} {\dot x}^\mu {\dot x}^\nu$ one finds that the one-graviton exchange diagrams vanish identically while the diagrams from Fig.~\ref{fig:v6logv2} cancel, so that the two-body potential is independent of the coefficients $c^{(a)}_V$.  We have dropped parts of diagrams that give rise to a contact delta function potential that does not generate long range forces, as well as terms that can be expressed as a total time derivative and therefore do not contribute to the equations of motion.  Note in particular that had we not chosen the subtraction scale $\mu$ to cancel the logarithms coming from the divergent integrals we would have needed, in order to obtain Eq.~(\ref{eq:v6logpot}), a contribution from a diagram involving the quintic vertex in $H_{\mu\nu}$ from the Einstein-Hilbert Lagrangian. 
 By choosing $\mu\sim 1/|{\bf x}_{12}|$, this piece of the potential arises instead from the diagram in Fig.~\ref{fig:v6logv2}f, which is much easier to calculate\footnote{This is true at least when the RG equations Eq.~(\ref{eq:RGcR}), Eq.~(\ref{eq:RGcV}) are calculated in background field gauge, as in the previous section.}.  To see this note that the insertion of the part of the operator $\int d\tau R$ that is quadratic in $H_{{\bf k}\mu\nu}$ in the diagram of Fig.~\ref{fig:v6logv2}f is related by gauge invariance to the short distance piece (the divergent log) that is generated in Fig.~\ref{fig:tadpoles}.

Defining the binding energy of the binary system in terms of the two-body Lagrangian as $E=\sum_a {\bf v}_a \cdot (\partial L/\partial {\bf v}_a) - L$ we find for a circular orbit of radius $r$ in the CM frame $\sum_a m_a {\bf x}_a=0$,
\begin{equation}
\label{eq:v6energy}
E_{v^6\ln\alpha}={1\over 3} \left({G_N m\over r}\right)^4 \mu\left(1-{3\mu\over m}\right)\ln {r\over r_0},
\end{equation}
where $\mu=m_1 m_2/m$, $m=m_1+m_2$, and $r_0$ is a length scale of order the radii of the binary star constituents, which we assume for simplicity to be of similar size.  One can check that this result agrees with the energy of a test particle of mass $\mu$ in circular motion around a Schwarzschild black hole of mass $m$ (in the coordinates used in Section~\ref{sec:RG}) in the limit $\mu\rightarrow 0$.  This requires knowledge of the logarithmic term in $g_{00}(r)$ proportional to $1/r^4,$ $g_{00}=1+\cdots + {4\over 3} (G_N m/r)^4\ln \mu r$.  

Because $r$ is not a physical quantity, it is more natural to express the binding energy $E$ in terms of the orbital frequency $\omega,$ which can be measured by observers at infinity.   A short calculation
shows that the contribution to $E(\omega)$ due to the coefficients $c_{R,V}(\mu)$ exactly cancels, which by RG invariance also implies that the logarithmic dependence on $\mu$ must disappear.  This follows from the metric field redefinitions alluded to earlier.  If we write the metric $g_{\mu\nu}$ in terms of a new metric ${\bar g}_{\mu\nu}$ as
\begin{equation}
\label{eq:fredef}
g_{\mu\nu}(x)={\bar g}_{\mu\nu}(x)\left[1 + {\xi\over 2 m^2_{Pl}} \int d\tau {\delta^4(x-x(\tau))\over\sqrt{\bar g}}\right],
\end{equation} 
we find that to linear order in $\xi$
\begin{equation}
-2 m^2_{Pl}\int d^4 x \sqrt{g} R[g]=-2m^2_{Pl}\int d^4 x \sqrt{{\bar g}} R[{\bar g}] +\xi\int d\tau R,
\end{equation}
so that the theory with curvature coupling $c_R$ is equivalent to a theory with coupling $c_R+\xi$.  Since $\xi$ can be chosen arbitrarily without affecting physical predictions, it can be set to $\xi=-c_R$ in which case the effects of the operator $\int d\tau R$ are completely removable\footnote{Note however, that the shift in Eq.~(\ref{eq:fredef})  induces in the two-body sector contact interactions of the form $\int d\tau\delta^4(x_1(\tau)-x_2(\tau))/\sqrt{g}$ which are physical, although irrelevant in classical processes.}.  Similar field redefinitions can be performed to remove $c_V$.   This conclusion can also be reached at the level of Eq.~(\ref{eq:v6logpot}).   Since to derive Eq.~(\ref{eq:v6logpot}) we had to fix a choice of gauge, a shift in the metric becomes equivalent to a gauge transformation.  It is then easy to see that if we let ${\bf x}_a\rightarrow {\bf x}_a+\delta{\bf x}_a$, ($a=1,2$) with $\delta{\bf x}_1=- 4 G_N c^{(2)}_R(\mu) {\bf x}_{12}/|{\bf x}_{12}|^3$  and  $\delta{\bf x}_2= 4 G_N c^{(1)}_R(\mu) {\bf x}_{12}/|{\bf x}_{12}|^3,$  the shift in the leading order Newtonian Lagrangian cancels the $v^6$ potentials calculated above. 

Beyond these spurious operators, any other operator involving the Ricci tensor is also removable by suitable metric redefinitions.   This means that only operators built from the Weyl tensor have any physical content.   The simplest physical operators that can be written down in a theory of spinless particles are then
\begin{eqnarray}
\nonumber
&\int d\tau C_{\mu\nu\alpha\beta} C^{\mu\nu\alpha\beta}, \int d\tau {\dot x}^\rho {\dot x}_\sigma C_{\mu\nu\alpha\rho}  C^{\mu\nu\alpha\sigma}&,\\
&\int d\tau {\dot x}^\alpha  {\dot x}^\beta {\dot x}_\rho  {\dot x}_\sigma C_{\mu\alpha\nu\beta} C^{\mu\rho\nu\sigma},
\end{eqnarray}
where we did not build any operators using the four-acceleration $a^\mu$, since such terms can be set to zero by a redefinition of the point particle worldline.  The NRGR power counting rules then indicate that these operators do not give finite size contributions to the gravitational wave observables until at least order $v^{10}$.   This is consistent with Newtonian expectations.   A complete account of finite size effects in the point particle EFT must also involve a discussion of absorptive effects, which for spinless black hole binaries in the extreme mass limit are known to arise first at order $v^8$~\cite{poisson}.  Such effects are also encoded in point particle operators, although there are subtleties involved in the construction which we will discuss in a separate paper.

\section{Conclusions}

In this paper, we have reformulated the NR expansion for binary star systems, an important class of signals for the gravitational wave detection experiments, in the language of effective field theory.  The EFT approach is useful for the binary problem because it allows the different length scales that arise in the binary star to be treated independently.  In particular, it is possible to separate the model dependent aspects of the problem which involve the internal structure of the binary constituents from the gravitational dynamics that governs the orbital evolution and radiation emission.  Using the EFT formulation, it is easy to show that the divergences which arise at $v^6$ in the PN expansion can be attributed to the existence of new operators in the effective point particle description. However, these operators can be removed via a point transformation of the metric tensor and thus never contribute to physical quantities. This leads to the conclusion that there are no finite size effects at order $v^6$.  Practically, this means that {\it whenever one encounters a log divergent integral at order $v^6$ in
the potential, one may simply set it to zero.  Its value cannot affect physical predictions.}

 Beyond order $v^6$, a new set of operators, which are quadratic in the curvature and whose role is to describe tidal deformations (induced moments), contribute at $v^{10}$.  The magnitude of such tidal forces is consistent with conventional Newtonian estimates of their size.  However, in the EFT these conclusions follow from general principles (dimensional analysis and general covariance) without the need to carry out any calculations.   As will be discussed in a future  publication, the calculations necessary to determine the coefficient of these operators (which encode the finite size effects) are relatively straightforward, since they can be done for a compact object in isolation without the need to include superfluous information about the complete two-body problem.  This is the beauty of working within an EFT formalism.

Several aspects of the formalism remain to be worked out.    Here we have not discussed logarithmic divergences that may arise in NRGR due to graphs with internal radiation graviton lines.  Such divergences can be treated by the same renormalization and matching procedure described in this paper to deal with divergences in the calculation of potential terms.  In fact, we believe that the explicit decoupling of the orbital scale $r$ and the radiation scale $r/v$ will make the evaluation of effects such as radiative tail terms, which is difficult within conventional PN approaches~\cite{PNv3rad,PNv4rad}, simpler within our approach.  Also, in our analysis so far, we have neglected the role of spins or higher rank tensors carried by the point particles.  Such permanent moments encode the intrinsic, rather than the tidally induced, shape of the binary star constituents.  It is important to extend the formalism, in particular the velocity power counting rules, to systematically include the effects of these additional worldline degrees of freedom on the pattern of gravitational waves seen by the detector.  We expect also that the inclusion of these dynamical quantities will allow the construction of a vastly larger set of coordinate invariant worldline non-minimal couplings than those considered in this paper, and lead to an even richer structure of coupling constant RG flows.  Finally it should be interesting to apply the formalism to work out the phenomenological signatures of theories of gravity that predict deviations from general relativity, including simple extensions with additional fields or more speculative models that deviate from conventional gravity in the far infrared~\cite{DGP,ghost}.

\acknowledgments

We thank Aneesh Manohar and Kip Thorne for helpful discussions.   WG is supported in part by the Department of Energy under Grant DE-FG03-97ER40546.  WG carried out part of this research as a postdoctoral fellow at UC Berkeley and thanks the financial support of the Berkeley theory group.  IZR would  like to thank the theory groups at UCSD and UC Berkeley for their hospitality.  IZR is supported in part by the Department of Energy under Grants DOE-ER-40682-143 and DEAC02-6CH03000.

{\bf Note added:}  While this paper was being reposted, Ref.~\cite{bdei} appeared which discusses the same field redefinitions for removing the operators $\int d\tau R$, $\int d\tau v^\mu v^\nu R_{\mu\nu}$ mentioned above.

\appendix
\section{Power counting $L$}

In this appendix we show that any diagram contributing to $S_{eff}[x_a]$ scales at most as $L^1$, up to powers of the velocity.  Consider a vacuum diagram with $N_g$ graviton self-interaction vertices from the Einstein-Hilbert action, $N_m$ insertions of the point particle worldline, and $P$ graviton propagators.  Writing the graviton kinetic term as $\sim m^2_{Pl}\int d^4 x (\partial h)^2$, each graviton self-interaction scales as $m^2_{Pl}$, a coupling to the worldlines scales as the particle mass $m$, and each graviton propagator scales as $1/m^2_{Pl}$.  Up to powers of $v$, the only other scale that can appear in a Feynman diagram is the orbital distance $r$.  Thus a typical diagram in the expansion of $S_{eff}[x_a]$ has a magnitude
\begin{equation}
m^{N_m} m^{2N_g}_{Pl} m^{-2P}_{Pl} r^{N_m+2N_g-2P},
\end{equation}
since $S_{eff}[x_a]$ is dimensionless.  Using the virial theorem, $m^{-2}_{Pl}\sim v^2 r/m$, so up to powers of $v$, this equation implies that each diagram scales as $(m r)^{N_m+N_g-P} \sim L^{V-P},$ where $V=N_m+N_g$ is the total number of vertices in each connected vacuum diagram that contributes to $S_{eff}[x_a]$.  

Starting with the trivial diagram with $V=1,P=0$ (corresponding to a free worldline) one can attach to it enough vertices and graviton lines to build up a diagram with arbitrary $V,P$.  However, since the diagram obtained in this way must be connected, for each new vertex we must include at least one graviton line, so that $V-P$ for this new graph is bounded by unity.  From this we conclude that any diagram contributing to $S_{eff}[x_a]$ scales as $L^{1-\ell}$, with integer $\ell\geq 0.$

\section{Feynman rules}

In this appendix we review the derivation of the Feynman rules used to obtain the results of this paper.  
As explained in the text, in our formalism all observables are derivable from either the quantity $S_{eff}[x_a]$ defined by Eq.~(\ref{eq:ZOSOPI}) or from the effective Lagrangian for the radiation graviton mode, $S_{NRGR}[{\bar h},x_a]$.  Note that because there is no propagator for the ``fields" $x^\mu_a(\tau)$ (as far as the gravitons are concerned, these degrees of freedom are treated as a fixed background) we have the rule 
\begin{itemize}
\item $iS_{eff}[x_a]$ is the sum of Feynman diagrams that remain connected when all particle worldlines are removed.   
\end{itemize}  
Thus diagrams like the Newton exchange term in Fig.~\ref{fig:newt}a are kept in the expansion, but diagrams such as a ``box" diagram with a pair of worldlines interchanging two gravitons do not arise in the expansion of $S_{eff}[x_a]$.  In fact, in our formalism the box diagram corresponds to the product of two single exchange terms, and is already accounted for by keeping just Fig.~\ref{fig:newt}a.

The procedure for integrating out modes with wavelengths shorter than the radiation scale $r/v$ involves computing diagrams with one or more external radiation gravitons and any number of internal potential gravitons.   This gives rise to the quantity $S_{NRGR}[{\bar h}, x_a]$ defined in Eq.~(\ref{eq:NRGRPI}).  The rule for calculating this is
\begin{itemize}
\item
A term with $n$ powers of the radiation field ${\bar h}_{\mu\nu}$ in $i S_{NRGR}[{\bar h},x_a]$ is equal to the sum of Feynman diagrams with $n$ external radiation graviton lines.
\end{itemize}
Due to the multipole expansion, this quantity is guaranteed to be of the general form $S_{NRGR}[{\bar h},{\bf x}_a]=\int dt L[{\bar h}({\bf X}_{cm},t),{\bf x}_a]$, as we explicitly saw in the derivation of the low order terms in $S_{eff}[{\bar h},x_a]$.

We now give the Feynman rules for computing the diagrams in the expansion of $iS_{eff}[x_a]$ and $i S_{NRGR}[{\bar h},x_a]$.

\subsection{Propagators and external gravitons}

If we expand the metric perturbation $h_{\mu\nu}=g_{\mu\nu}-\eta_{\mu\nu}$ in terms of potential and radiation graviton modes as
\begin{equation}
h_{\mu\nu}(x)={\bar h}_{\mu\nu}(x) +\int_{\bf k} {H_{\bf k}}_{\mu\nu}(x^0) e^{i{\bf k}\cdot {\bf x}}
\end{equation}
then the propagators for ${H_{\bf k}}_{\mu\nu}$ and ${\bar h}_{\mu\nu}$ can be read off the terms in the gravitational action that are quadratic in the fields.   In order to get a well defined propagator one needs to fix the gauge, which is achieved by adding a term to the action that explicitly breaks diffeomorphisms This procedure is standard for gauge theories and can be found for instance in chapter 16 of Ref.~\cite{pands}.  For gravity this is explained in the reviews~\cite{GREFT,GREFTrev}.  Given the choice of gauge fixing term for potential gravitons used in the text, Eq.~(\ref{eq:pgauge})  the terms quadratic in the potential mode can be written as
\begin{equation}
S_{H^2}= -{1\over 2} \int dx^0\int_{\bf k} {H_{-{\bf k}}}_{\mu\nu} (x^0){\bf k}^2 T^{\mu\nu;\alpha\beta} {H_{{\bf k}}}_{\alpha\beta}(x^0), 
\end{equation}
with $T^{\mu\nu;\alpha\beta}={1\over 2}\eta^{\mu\alpha}\eta^{\nu\beta}+{1\over2}\eta^{\mu\beta}\eta^{\nu\alpha}-{1\over 2}\eta^{\mu\nu}\eta^{\alpha\beta}$.  The propagator is then defined as a formal inverse inverse to this equation
\begin{eqnarray}
\label{eq:propeqn}
\nonumber
i {\bf k}^2 T^{\mu\nu;\rho\sigma}\langle {H_{\bf k}}_{\rho\sigma}(x^0)  {H_{\bf q}}_{\alpha\beta}(x'^0) \rangle = \\
(2\pi)^3 \delta^3({\bf k}+{\bf q}) \delta(x^0-x'^0) I^{\mu\nu}_{\alpha\beta},
\end{eqnarray}
with $I^{\mu\nu;\alpha\beta}={1\over 2}\eta^{\mu\alpha}\eta^{\nu\beta}+{1\over 2}\eta^{\mu\beta}\eta^{\nu\alpha}$ the identity on symmetric two-index tensors.   Inverting we find Eq.~(\ref{eq:potprop})
\begin{equation}
\label{eq:pprop}
\langle {H_{\bf k}}_{\mu\nu}(x^0)  {H_{\bf q}}_{\alpha\beta}(0) \rangle =  -(2\pi)^3\delta^3({\bf k}+{\bf q}){i\over {\bf k}^2}\delta(x^0) P_{\mu\nu;\alpha\beta}.
\end{equation}
The propagator for the radiation graviton also depends on the choice of gauge fixing term in the action.   For instance, choosing a gauge fixing term of the form
\begin{equation}
\label{eq:rgauge}
S_{GF}[{\bar h}]= m^2_{Pl}\int d^d x \eta^{\mu\nu} {\bar\Gamma}_\mu {\bar\Gamma}_\nu,
\end{equation}
with ${\bar\Gamma}_\mu =\partial_\alpha {{\bar h}^\alpha}_{\mu} -{1\over 2}\partial_\mu {\bar h},$ the propagator is given by 
\begin{equation}
\label{eq:rprop}
\langle {\bar h}_{\mu\nu}(x) {\bar h}_{\alpha\beta}(y) \rangle = D_F(x-y) P_{\mu\nu;\alpha\beta},
\end{equation}
where the Feynman propagator $D_F(x)$ is given by
\begin{equation}
D_F(x)=\int {d^d k\over (2\pi)^d} {i\over k^2+i\epsilon} e^{-i k\cdot x}.
\end{equation}
In this equation,  $\epsilon\rightarrow 0^+$ defines a choice of contour for the  $k^0$ integral.    Given these results, the general rule for a given Feynman diagram is then:
\begin{itemize}
\item  An internal potential graviton line with label momentum ${\bf k}$ and connecting points at times $x^0$ and $x'^0$ corresponds to an insertion of Eq.~(\ref{eq:pprop}).
\item 
An internal radiation line connecting two points $x$ and $y$ denotes a factor of Eq.~(\ref{eq:rprop}).  
\end{itemize}

\begin{itemize}
\item For each external radiation line starting at a vertex located at a spacetime point  $x$, one includes a factor of the radiation field ${\bar h}_{\mu\nu}(x)$ and an integral $\int d^d x$ over $x$.   
\end{itemize}

\subsection{Vertices}

\begin{figure}
\label{fig:h00feyn}
    \centering
    \includegraphics[width=4cm]{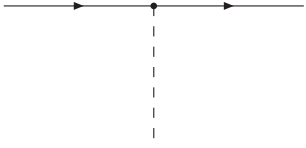}
\caption{Vertex associated with the Feynman rule in Eq.~(\ref{eq:h00feyn})}
\end{figure}

They Feynman rules for general relativity contain an infinite number of interaction vertices.   This is because the Einstein-Hilbert Lagrangian $\sqrt{g} R$ is an infinite series in the graviton $h_{\mu\nu}$, so that the set of terms in this expansion with $n$ factors of $h_{\mu\nu}$ gives rise to an $n$-graviton vertex in Feynman diagrams.  In our theory, the expansion of the point particle couplings also generates an infinite set of vertices that can be used to build up Feynman diagrams.   The method for deriving the vertex rules associated with these interactions is straightforward, although in the case of graviton self-interactions arising from the gravitational action the procedure gets messy due to the large number of terms at a given order in $h_{\mu\nu}$.  In this section we will illustrate the procedure by deriving some of the vertex rules used in the text.  The derivation of more complicated Feynman rules is directly analogous.  

Consider first Feynman rules arising from insertions of point particle interactions.  The simplest non-trivial term is the vertex that arises in the calculation of the two-body Newton interaction.   Expanding the $-m\int d\tau$ term in the action, there is a term of the form
\begin{equation}
S_{pp}=-{m\over 2 m_{Pl}}\int dx^0\int_{\bf k} e^{i{\bf k}\cdot {\bf x}(x^0)} H_{{\bf k}00}(x^0)+\cdots.
\end{equation}
This leads to a Feynman rule for the vertex shown in Fig.~11
\begin{equation}
\label{eq:h00feyn}
-{i m\over 2 m_{Pl}}\int dx^0 \int_{\bf k} e^{i{\bf k}\cdot {\bf x}(x^0)}\eta_{\mu0}\eta_{\nu0}.
\end{equation}
Then for instance the diagram of Fig.~\ref{fig:newt}a is given by 
\begin{widetext}
\begin{equation}
\mbox{Fig.~\ref{fig:newt}a} =\left[-{i m_1\over 2 m_{Pl}}\int dx^0 \int_{\bf k} e^{i{\bf k}\cdot {\bf x}_1(x^0)}\eta_{\mu0}\eta_{\nu0}\right] \left[-{i m_2\over 2 m_{Pl}}\int dx'^0 \int_{\bf q} e^{i{\bf q}\cdot {\bf x}_2(x'^0)}\eta_{\alpha0}\eta_{\beta0}\right] \langle H^{\mu\nu}_{\bf k}(x^0) H^{\alpha\beta}_{\bf q}(x'^0)\rangle,
\end{equation}
\end{widetext}
which reduces to Eq.~(\ref{eq:newta}).  Likewise, the term in the action
\begin{equation}
S_{pp}=-{m\over 2 m_{Pl}}\int dx^0\int_{\bf k} e^{i{\bf k}\cdot{\bf x}(x^0)} {\bf v}_i(x^0) H_{{\bf k}0i}(x^0)+\cdots,
\end{equation}
leads to the Feynman rule
\begin{equation}
-{i m\over 2 m_{Pl}} \int dx^0\int_{\bf k} e^{i{\bf k}\cdot{\bf x}(x^0)} {\bf v}_i(x^0) \delta_{i\mu}\eta_{\nu0},
\end{equation}
which can be used to calculate Fig.~\ref{fig:EIH1}b.  The point particle action has also terms that are non-linear in the graviton and lead to vertices with more graviton lines.   For instance, the vertex in Fig.~\ref{fig:EIH2}b  involving two graviton lines arises from a term in the action
\begin{equation}
S_{pp}={m\over 8 m^2_{Pl}}\int dx^0\int_{{\bf k},{\bf q}} e^{i({\bf q}+{\bf k})\cdot {\bf x}(x^0)} H_{{\bf k}00}(x^0) H_{{\bf q}00}(x^0).
\end{equation}
The Feynman rule for this term contains two identical terms, which arise from the two independent ways of contracting the two fields in the interaction with the external graviton lines (in other words, of pairing the two graviton lines with index pairs $\mu\nu$, $\alpha\beta$ with the two fields in the interaction)
\begin{equation}
\label{eq:seag}
{i m\over 4 m^2_{Pl}}\int dx^0 \int_{{\bf k},{\bf q}} e^{i({\bf q}+{\bf k})\cdot {\bf x}(x^0)} \eta_{0\mu}\eta_{0\nu}\times\eta_{0\alpha}\eta_{0\beta}.
\end{equation}
Using this vertex, we can then calculate the graph in Fig.~\ref{fig:EIH2}b.  Including the proper symmetry factors for the diagram (see for instance chapter 4 of Ref.~\cite{pands}) leads to Eq.~(\ref{eq:EIH5}).

Besides the interactions vertices involving point particle sources, we also need to calculate the $n$-graviton vertices.  Consider for instance the 3-point vertex for radiation gravitons.   Due to the large number of terms each with a complicated tensor index structure, we do not display the full Feynman rule here (it is more efficient to do the calculations using an algebra manipulation computer package).   Consider, however one of the terms that gives rise to the 3-graviton Feynman rule.  In the gauge of Eq.~(\ref{eq:rgauge}) all terms in this vertex derive from the cubic terms in the expansion of $\sqrt{g} R$.  For instance, using 
\begin{eqnarray}
\nonumber
\sqrt{g} g^{\mu\nu} &=& \eta^{\mu\nu} +\cdots+ {1\over 4 m^2_{Pl}}\left[{\bar h}^2 -{1\over 2} {\bar h}_{\alpha\beta} {\bar h}^{\alpha\beta}\right]\eta^{\mu\nu}\\
& &  -{1\over 2 m^2_{Pl}} {\bar h} {\bar h}^{\mu\nu}+\cdots,
\end{eqnarray}
there is a term in the Einstein-Hilbert Lagrangian of the form
\begin{widetext}
\begin{equation}
\label{eq:part}
-2m^2_{Pl}\int d^d x \sqrt{g} R = - {1\over 2 m_{Pl}}\int d^4 x \left[{\bar h}^2 -{1\over 2} {\bar h}_{\alpha\beta} {\bar h}^{\alpha\beta}\right] R^{(1)} + {1\over m_{Pl}}\int d^4 x {\bar h} {\bar h}^{\mu\nu} R^{(1)}_{\mu\nu}+\cdots,
\end{equation}
\end{widetext}
where $R^{(1)}_{\mu\nu}={1\over 2}\partial_\nu \partial_{\alpha}{\bar h}^{\alpha}_\mu+{1\over 2}\partial_\mu\partial_{\alpha}{\bar h}^{\alpha}_\nu-{1\over2}\partial^2{\bar h}_{\mu\nu}-{1\over 2}\partial_\mu \partial_\nu {\bar h}$, and $R^{(1)}=\eta^{\mu\nu} R^{(1)}_{\mu\nu}$.   In coordinate space this particular term leads to a contribution to the 3-graviton vertex of the form
\begin{widetext}
\begin{eqnarray}
\nonumber
{i\over m_{Pl}}\int d^d x \left[\left(-{1\over 2}  {{I_{\mu_1\nu_1}}^{\alpha}}_{\alpha}  {{I_{\mu_2\nu_2}}^{\beta}}_{\beta}  + {1\over 4} I_{{\mu}_1{\nu_1}\alpha\beta}  I^{\alpha\beta}_{{\mu}_2{\nu_2}}\right)I_{\mu_3\nu_3\rho\sigma}\left(\partial_3^\rho \partial_3^\sigma-\eta^{\rho\sigma}\partial_3^2\right)\right.\\
+  \left. {{I_{\mu_1\nu_1}}^{\rho}}_{\rho} I_{\mu_2\nu_2\alpha\beta}\left({1\over 2}{I^{\sigma\beta}}_{\mu_3\nu_3} {\partial_3}^\alpha {\partial_3}_\sigma +{1\over 2}{I^{\sigma\alpha}}_{\mu_3\nu_3} {\partial_3}^\beta {\partial_3}_\sigma -{1\over 2} {I^{\alpha\beta}}_{\mu_3\nu_3}\partial_3^2 - {1\over 2} {{I_{\mu_3\nu_3}}^{\sigma}}_{\sigma}  \partial^\alpha_3\partial^\beta_3\right)\right]+\mbox{perms}.
\end{eqnarray}
\end{widetext}
The tensor $I_{\mu\nu\alpha\beta}$ has been defined below Eq.~(\ref{eq:propeqn}).  In this vertex, each incoming graviton is labeled 1, 2, or 3.   The partial derivatives in this particular term act on the propagator of the graviton labeled 3 coming into the vertex.   To get the full Feynman rule from the term in Eq.~(\ref{eq:part}), Bose statistics also demands that we include all permutations of the terms shown in the above expression under the interchange of the graviton labels.  It is also useful to work in momentum space, in which case the expression for this piece of the 3-graviton Feynman rule becomes (taking all graviton momenta to be flowing into the vertex)
\begin{widetext}
\begin{eqnarray}
\nonumber
-{i\over m_{Pl}}\ (2\pi)^d \delta^d(k_1+k_2+k_3)\left[\left(-{1\over 2}  {{I_{\mu_1\nu_1}}^{\alpha}}_{\alpha}  {{I_{\mu_2\nu_2}}^{\beta}}_{\beta}  + {1\over 4} I_{{\mu}_1{\nu_1}\alpha\beta}  I^{\alpha\beta}_{{\mu}_2{\nu_2}}\right)I_{\mu_3\nu_3\rho\sigma}\left(k_3^\rho k_3^\sigma-\eta^{\rho\sigma} k_3^2\right)\right.\\
+  \left. {{I_{\mu_1\nu_1}}^{\rho}}_{\rho} I_{\mu_2\nu_2\alpha\beta}\left({1\over 2}{I^{\sigma\beta}}_{\mu_3\nu_3} {k_3}^\alpha {k_3}_\sigma +{1\over 2}{I^{\sigma\alpha}}_{\mu_3\nu_3} {k_3}^\beta {k_3}_\sigma -{1\over 2} {I^{\alpha\beta}}_{\mu_3\nu_3}k_3^2 - {1\over 2} {{I_{\mu_3\nu_3}}^{\sigma}}_{\sigma}  k^\alpha_3 k^\beta_3\right)\right]+\mbox{perms}.
\end{eqnarray}
\end{widetext}
Feynman rules for potential self-couplings, or for potential-radiation couplings are derived in the same way.


\begin{references}



\bibitem{LIGO}
A.~Abramovici {\it et al.},
Science {\bf 256}, 325 (1992);

\bibitem{VIRGO}
A.~Giazotto,
Nucl.\ Instrum.\ Meth.\ A {\bf 289}, 518 (1990).


\bibitem{EIH} 
A.~Einstein, L.~Infeld and B.~Hoffmann,
Annals Math.\  {\bf 39}, 65 (1938).

\bibitem{PNformalism}
R.~V.~Wagoner and C.~M.~Will,
Astrophys.\ J.\  {\bf 210}, 764 (1976)
[Erratum-ibid.\  {\bf 215}, 984 (1977)];
M.~Walker and C.~M.~Will,
Phys.\ Rev.\ Lett.\  {\bf 45}, 1741 (1980);
L.~Blanchet, T.~Damour and G.~Schafer,
Mon.\ Not.\ Roy.\ Astron.\ Soc.\  {\bf 242}, 289 (1990).



\bibitem{dataanalysis}
C.~Cutler {\it et al.},
Phys.\ Rev.\ Lett.\  {\bf 70}, 2984 (1993)
[arXiv:astro-ph/9208005];
C.~Cutler and E.~E.~Flanagan,
Phys.\ Rev.\ D {\bf 49}, 2658 (1994)
[arXiv:gr-qc/9402014];
L.~S.~Finn,
Phys.\ Rev.\ D {\bf 46}, 5236 (1992)
[arXiv:gr-qc/9209010];
L.~S.~Finn and D.~F.~Chernoff,
Phys.\ Rev.\ D {\bf 47}, 2198 (1993)
[arXiv:gr-qc/9301003];
E.~Poisson and C.~M.~Will,
Phys.\ Rev.\ D {\bf 52}, 848 (1995)
[arXiv:gr-qc/9502040].

\bibitem{PNv3rad}
A.~G.~Wiseman,
Phys.\ Rev.\ D {\bf 48}, 4757 (1993);
L. Blanchet and G. Schafer, Class. Quant. Grav.~{\bf 10}, 2699 (1993);
L.~Blanchet and T.~Damour,
Phys.\ Rev.\ D {\bf 46}, 4304 (1992).


\bibitem{PNv4pot}
T.~Damour and N.~Deruelle,
Phys.\ Lett.\ A {\bf 87}, 81 (1981);
C.R. Acad. Sc. Paris {\bf 293}, 537 (1981); 
C.R. Acad. Sc. Paris {\bf 294}, 1355 (1982);
L. Bel, T. Damour, N. Deroulle, J. Ibanez, and J. Martin, Gen. Rel. Grav. {\bf 13}, 963 (1981);
T. Damour and G. Schafer, Gen. Rel. Grav. {\bf 17}, 879 (1985).
 
 \bibitem{PNv4rad}
L .~Blanchet, T.~Damour and B.~R.~Iyer,
Phys.\ Rev.\ D {\bf 51}, 5360 (1995)
[Erratum-ibid.\ D {\bf 54}, 1860 (1996)]
[arXiv:gr-qc/9501029];
C.~M.~Will and A.~G.~Wiseman,
Phys.\ Rev.\ D {\bf 54}, 4813 (1996)
[arXiv:gr-qc/9608012];
L.~Blanchet, B.~R.~Iyer, C.~M.~Will and A.~G.~Wiseman,
Phys.\ Rev.\ Lett.\  {\bf 74}, 3515 (1995)
[arXiv:gr-qc/9501027];
Class.\ Quant.\ Grav.\  {\bf 13}, 575 (1996)
[arXiv:gr-qc/9602024].

\bibitem{PNv5pot}
S.M. Kopejkin, Astron. Zh. {\bf 62}, 889 (1985); 
L.~Blanchet, G.~Faye and B.~Ponsot,
Phys.\ Rev.\ D {\bf 58}, 124002 (1998)
[arXiv:gr-qc/9804079];
M.~E.~Pati and C.~M.~Will,
Phys.\ Rev.\ D {\bf 62}, 124015 (2000)
[arXiv:gr-qc/0007087];
Phys.\ Rev.\ D {\bf 65}, 104008 (2002)
[arXiv:gr-qc/0201001];
Y.~Itoh, T.~Futamase and H.~Asada,
Phys.\ Rev.\ D {\bf 62}, 064002 (2000)
[arXiv:gr-qc/9910052];
Phys.\ Rev.\ D {\bf 63}, 064038 (2001)
[arXiv:gr-qc/0101114];

\bibitem{PNv56rad}
L.~Blanchet,
Phys.\ Rev.\ D {\bf 54}, 1417 (1996)
[arXiv:gr-qc/9603048];
L.~Blanchet, B.~R.~Iyer and B.~Joguet,
Phys.\ Rev.\ D {\bf 65}, 064005 (2002)
[arXiv:gr-qc/0105098];
L.~Blanchet, G.~Faye, B.~R.~Iyer and B.~Joguet,
Phys.\ Rev.\ D {\bf 65}, 061501 (2002)
[arXiv:gr-qc/0105099].


\bibitem{PNv6pot}
P.~Jaranowski and G.~Schafer,
Phys.\ Rev.\ D {\bf 57}, 7274 (1998)
[Erratum-ibid.\ D {\bf 63}, 029902 (2001)]
[arXiv:gr-qc/9712075];
Phys.\ Rev.\ D {\bf 60}, 124003 (1999)
[arXiv:gr-qc/9906092];
T.~Damour, P.~Jaranowski and G.~Schafer,
Phys.\ Rev.\ D {\bf 62}, 021501 (2000)
[Erratum-ibid.\ D {\bf 63}, 029903 (2001)]
[arXiv:gr-qc/0003051];
Phys.\ Rev.\ D {\bf 63}, 044023 (2001)
[arXiv:gr-qc/0010009];
L.~Blanchet and G.~Faye,
Phys.\ Lett.\ A {\bf 271}, 58 (2000);
[arXiv:gr-qc/0004009].
Phys.\ Rev.\ D {\bf 63}, 062005 (2001)
[arXiv:gr-qc/0007051].
V.~C.~de Andrade, L.~Blanchet and G.~Faye,
Class.\ Quant.\ Grav.\  {\bf 18}, 753 (2001)
[arXiv:gr-qc/0011063].
L.~Blanchet and B.~R.~Iyer,
Class.\ Quant.\ Grav.\  {\bf 20}, 755 (2003)
[arXiv:gr-qc/0209089].

\bibitem{PNambiguity}
T.~Damour, P.~Jaranowski and G.~Schafer,
Phys.\ Rev.\ D {\bf 63}, 044021 (2001)
[Erratum-ibid.\ D {\bf 66}, 029901 (2002)]
[arXiv:gr-qc/0010040];
T.~Damour, P.~Jaranowski and G.~Schafer,
Phys.\ Rev.\ D {\bf 62}, 021501 (2000)
[Erratum-ibid.\ D {\bf 63}, 029903 (2001)]
[arXiv:gr-qc/0003051];
L.~Blanchet, T.~Damour, G.~Esposito-Farese and B.~R.~Iyer,
arXiv:gr-qc/0406012.

\bibitem{PNdimreg}
T.~Damour, P.~Jaranowski and G.~Schafer,
Phys.\ Lett.\ B {\bf 513}, 147 (2001)
[arXiv:gr-qc/0105038].
L.~Blanchet, T.~Damour and G.~Esposito-Farese,
Phys.\ Rev.\ D {\bf 69}, 124007 (2004)
[arXiv:gr-qc/0311052];

\bibitem{NREFT}
W.~E.~Caswell and G.~P.~Lepage,
Phys.\ Lett.\ B {\bf 167}, 437 (1986);

\bibitem{LMR} M.~E.~Luke, A.~V.~Manohar and I.~Z.~Rothstein,
Phys.\ Rev.\ D {\bf 61}, 074025 (2000)
[arXiv:hep-ph/9910209].

\bibitem{EFTrev}
For a recent review, see I.~Z.~Rothstein,
arXiv:hep-ph/0308266.

\bibitem{poisson}
E.~Poisson and M.~Sasaki,
  Phys.\ Rev.\ D {\bf 51}, 5753 (1995)
  [arXiv:gr-qc/9412027].


\bibitem{will}  
T.~Mora and C.~M.~Will,
Phys.\ Rev.\ D {\bf 69}, 104021 (2004)
[arXiv:gr-qc/0312082].

\bibitem{us} W. Goldberger and I. Rothstein, in preparation.

\bibitem{HQET}
N.~Isgur and M.~B.~Wise,
Phys.\ Lett.\ B {\bf 232}, 113 (1989);
Phys.\ Lett.\ B {\bf 237}, 527 (1990);
E.~Eichten and B.~Hill,
Phys.\ Lett.\ B {\bf 234}, 511 (1990);
H.~Georgi,
Phys.\ Lett.\ B {\bf 240}, 447 (1990).


\bibitem{GREFT}
J.~F.~Donoghue,
Phys.\ Rev.\ Lett.\  {\bf 72}, 2996 (1994)
[arXiv:gr-qc/9310024];
J.~F.~Donoghue,
Phys.\ Rev.\ D {\bf 50}, 3874 (1994)
[arXiv:gr-qc/9405057];


\bibitem{GREFTrev} 
For reviews, see J.~F.~Donoghue,
arXiv:gr-qc/9512024;
C.~P.~Burgess,
arXiv:gr-qc/0311082.

\bibitem{GRNRPC}
J.~F.~Donoghue and T.~Torma,
Phys.\ Rev.\ D {\bf 54}, 4963 (1996)
[arXiv:hep-th/9602121].

\bibitem{de}
T.~Damour and G.~Esposito-Farese,
  Phys.\ Rev.\ D {\bf 58}, 042001 (1998)
  [arXiv:gr-qc/9803031].


\bibitem{hartlethorne}
K.~S.~Thorne and J.~B.~Hartle,
Phys.\ Rev.\ D {\bf 31}, 1815 (1984).

\bibitem{veltman}
M. Veltman, in \emph{Methods in Field Theory, Proceedings of the Les Houches Summer School, 1975,} eds. R. Balian and J. Zinn-Justin, North Holland, 1976.


\bibitem{QCDmult}
B.~Grinstein and I.~Z.~Rothstein,
Phys.\ Rev.\ D {\bf 57}, 78 (1998)
[arXiv:hep-ph/9703298].

\bibitem{worldline}
G.~Modanese,
Nucl.\ Phys.\ B {\bf 434}, 697 (1995)
[arXiv:hep-th/9408103];

\bibitem{muzinich}
I.~J.~Muzinich and S.~Vokos,
Phys.\ Rev.\ D {\bf 52}, 3472 (1995)
[arXiv:hep-th/9501083].


\bibitem{EIHquant}
N.~E.~J.~Bjerrum-Bohr, J.~F.~Donoghue and B.~R.~Holstein,
Phys.\ Rev.\ D {\bf 67}, 084033 (2003)
[arXiv:hep-th/0211072];
H.~W.~Hamber and S.~Liu,
Phys.\ Lett.\ B {\bf 357}, 51 (1995)
[arXiv:hep-th/9505182];
N.~E.~J.~Bjerrum-Bohr, J.~F.~Donoghue and B.~R.~Holstein,
Phys.\ Rev.\ D {\bf 68}, 084005 (2003)
[arXiv:hep-th/0211071].

\bibitem{QCDworldline}
W.~Fischler,
Nucl.\ Phys.\ B {\bf 129}, 157 (1977);
T.~Appelquist, M.~Dine and I.~J.~Muzinich,
Phys.\ Lett.\ B {\bf 69}, 231 (1977);
Phys.\ Rev.\ D {\bf 17}, 2074 (1978).

\bibitem{peskin}
M.~E.~Peskin,
Nucl.\ Phys.\ B {\bf 156}, 365 (1979);
for a review see M.~E.~Peskin,
SLAC-PUB-3273,
{\it Presented at 11th Int. SLAC Summer Inst. on Particle Physics, Stanford, CA, Jul 18-26, 1983.}


\bibitem{classicalRG}
S.~N.~Solodukhin,
Nucl.\ Phys.\ B {\bf 541}, 461 (1999)
[arXiv:hep-th/9801054];
W.~D.~Goldberger and M.~B.~Wise,
Phys.\ Rev.\ D {\bf 65}, 025011 (2002)
[arXiv:hep-th/0104170].

\bibitem{background}
B.~S.~Dewitt,
Phys.\ Rev.\  {\bf 162}, 1195 (1967);
L.~F.~Abbott,
Nucl.\ Phys.\ B {\bf 185}, 189 (1981).

\bibitem{DGP}
G.~R.~Dvali, G.~Gabadadze and M.~Porrati,
  Phys.\ Lett.\ B {\bf 485}, 208 (2000)
  [arXiv:hep-th/0005016].


\bibitem{ghost}
N.~Arkani-Hamed, H.~C.~Cheng, M.~A.~Luty and S.~Mukohyama,
JHEP {\bf 0405}, 074 (2004)
[arXiv:hep-th/0312099].

\bibitem{pands}
M. E. Peskin and D. V. Schroeder, \emph{An Introduction to Quantum Field Theory,} Perseus Books, 1995.

\bibitem{bdei}
L. Blanchet et. al, arXiv:gr-qc/0503044.


\end{references}
\end{document}